\documentclass[useAMS,usenatbib,subeqn]{mn2e}

\usepackage{multirow}  
\usepackage{color}
\usepackage{epsfig,amssymb}
\usepackage{graphicx,epsfig}
\usepackage{hhline}
\usepackage{subfigure}
\usepackage{eufrak}
\usepackage{mathrsfs}
\usepackage{eucal}
\usepackage{amsmath}
\usepackage{aas_macros}
\usepackage{fontenc}
\usepackage{textcomp}
\usepackage{natbib}

\newcommand{\fnl}{$f_{\mathrm{NL}}^{\mathrm{local}}\;$}

\title[Search for non-Gaussianity in the BOSS DR9 quasars]{Search for primordial non-Gaussianity in the quasars of SDSS-III BOSS DR9} \author[D. Karagiannis et al.] {D. Karagiannis$^{1}$\thanks{E-mail: dionysios.karagiannis@pd.infn.it},
T. Shanks$^{1}$, Nicholas P. Ross$^{2,3}$ \\ 
$^{1}$Physics Department, University of Durham, South Road, Durham, DH1 3LE, UK \\
$^{2}$Lawrence Berkeley National Laboratory, 1 Cyclotron Road, Berkeley, CA 94720, USA \\
$^{3}$Department of Physics, Drexel University, 3141 Chestnut Street, Philadelphia, PA 19104, USA }

\begin{document}

\pagerange{\pageref{firstpage}--\pageref{lastpage}} 

\maketitle 
\begin{abstract}

We analyse the clustering of 22,361 quasars between redshift $2.2<z<2.9$
observed with the Sloan Digital Sky Survey (SDSS)-III Baryon Oscillation
Spectroscopic Survey (BOSS), which are included in the ninth data release
(DR9). We fit the clustering results with a $\Lambda$CDM model to
calculate the linear bias of the quasar sample, \mbox{$b=3.74\pm0.12$}.
The measured value of bias is consistent with the findings of
\citet{white2012}, where they analyse almost the same quasar sample,
although only in the range, $\rm{s}<40\;h^{-1}\rm{Mpc}$. At large scales we
observe an excess or plateau in the clustering correlation function. By fitting a model that
incorporates a scale dependent additional term in the bias introduced by
primordial non-Gaussianity of the local type, we calculate the amplitude
of the deviation from the Gaussian initial conditions as
$70<f_{\mathrm{NL}}^{\mathrm{local}}<190$ at the $95\%$ confidence level. We correct the sample from systematics according to the methods of \citet{ross_2011,ho2012} and \citet{ross_2011,ross_2012}, with the \fnl
measurements after the application of the two methods being consistent
with each other. Finally we use cross-correlations across redshift slices to test the
corrected sample for any remaining unknown sources of  systematics, but the results give no indication of any such further errors. We consider as our final results on non-Gaussianity, $46<f_{\mathrm{NL}}^{\mathrm{local}}<158$ at $95\%$ confidence, after correcting the sample with the weights method of \citet{ross_2011,ross_2012}. These results are consistent with previous tight constraints on non-Gaussianity
from other LSS surveys, but are in tension with the latest results from the CMB.

\end{abstract}


\section{Introduction} 
Inflation is the leading scenario for describing the very early
Universe, solving at the same time major cosmological problems. One of
most the important aspects of the inflationary paradigm is the natural
production of primordial fluctuations in the density field, which will
constitute the origin of the observable structures. In the simplest
single field, slow roll inflationary model the \emph{inflaton} scalar
field, that drives the accelerating expansion of the Universe during
that era, can generate primordial curvature fluctuations. These
perturbations will create primordial gravitational potential and
eventually, through Poisson equation, density fluctuations. The inflaton
follows Gaussian statistics, as a quantum field, and therefore the
generated primordial perturbations in such models are also Gaussian.
However, observational data does not prevent a deviation from
Gaussianity. 

Many different types of inflation models violate one or more of the
simple inflation conditions and generate large non-Gaussianities.
Therefore in order to distinguish between all these different mechanisms
we have to gain additional information from the non-Gaussian part of the
primordial perturbations. The two-point correlation function can be used
to describe the density fluctuations in the Universe. Such statistics
describe Gaussian random fields, hence any information on
non-Gaussianity must be extracted from the higher order correlation
functions. The simplest and most studied non-Gaussian correlator is the
three-point correlation function and its Fourier coefficient the
\emph{bispectrum}. Their presence guarantees the departure from
Gaussianity.

    The presence of primordial non-Gaussianity of the local type affects the primordial gravitational potential perturbation, where by Taylor expanding around its Gaussian part we have \citep{gangui,verde_wang}

    \begin{equation} \label{eq:phi_ng}
    \Phi(\mathbf{x})=\Phi_G(\mathbf{x})+f^{\mathrm{local}}_{\mathrm{NL}}\left[\Phi_G^2(\mathbf{x})-\langle\Phi_G^2(\mathbf{x})\rangle\right]
    \end{equation}

    \noindent where $\Phi(\mathbf{x})$ is the Bardeen gauge-invariant potential in real space and $\Phi_G(\mathbf{x})$ is the Gaussian part of the potential. By using primordial here we mean the gravitational potential perturbations before the action of the transfer function. The \fnl parameter defines the deviation from the Gaussian initial conditions. Non-Gaussianity guarantees the presence of a gravitational potential bispectrum, which will be defined in the case of the local type as
    
    \begin{align}
 B_{\Phi}^{\mathrm{loc}}(\mathbf{k_1},\mathbf{k_2},\mathbf{k_3})=& 2f_{\mathrm{NL}}^{\mathrm{loc}}(P_{\phi}(\mathbf{k_1})P_{\phi}(\mathbf{k_2})+P_{\phi}(\mathbf{k_2})P_{\phi}(\mathbf{k_3}) \\ \nonumber
 &+P_{\phi}\mathbf{k_1})P_{\phi}(\mathbf{k_3}))
\end{align}

    \noindent where $B_{\Phi}^{\mathrm{loc}}(\mathbf{k_1},\mathbf{k_2},\mathbf{k_3})$ is the bispectrum of the primordial gravitational potential perturbation $\Phi(\mathbf{x})$ and $P_{\phi}(\mathbf{k})$\footnote{Here we used the convention $\phi\equiv\Phi_G$.} is the power spectrum of its Gaussian part $\Phi_G(\mathbf{x})$. It is clear that the amount of information bispectrum holds, in the case of a non-zero \fnl, is far greater than that of the power spectrum, which correlates only two points.

Mainly there are two ways to get information on non-Gaussianity in the
primordial perturbations, by measuring higher-order statistics (e.g.
bispectrum, trispectrum) of the CMB anisotropies and from the abundance
and clustering of the Large-Scale Structures (LSS). The CMB anisotropies
has provided the tighter constrains on the $f_{\mathrm{NL}}$ parameter
for the local regime, by directly measuring the bispectrum. More
precisely WMAP3 found $-36<f_{\mathrm{NL}}^{\mathrm{local}}< 100$ at
$95\%$ confidence level (CL) \citep{wmap3}, from WMAP5 
$-4<f_{\mathrm{NL}}^{\mathrm{local}}<80$ at $95\%$ CL
\citep{wmpa5_komatsu,wmap_smith}, WMAP7 measured
$f_{\mathrm{NL}}^{\mathrm{local}}=32\pm21$ at $68\%$ CL
\citep{komatsu_wmap7,wmap7_colab} and finally the recent results from
Planck measure $f_{\mathrm{NL}}^{\mathrm{local}}=2.7\pm5.8$ at $68\%$ CL
\citep{planck_2013_xxiv}. The exploitation of the CMB data provide
information on the cosmological fluctuations, and hence non-Gaussianity,
in their original primordial form.

Recently it has been found that the presence of local non-Gaussianity in
the initial perturbation density field can affect the dark matter halo
mass function \citep{MVJ,LMSV}, as well as the bias relationship of
galaxies and halos with the underlying matter distribution. It has been
proven \citep{dalal_2008,MV08} that an additional scale-dependent term
is introduced in the bias, $b_{NG}\sim k^{-2}$, where its amplitude is
quantified by \fnl. This result has been derived in many different
ways \citep{slosar_2008,afshordi_2008,mcdonald_2008} giving generally
the same answer. 

Taking advantage of this, we can measure any deviation from the Gaussian
conditions by fitting a non-Gaussian model in the clustering of galaxies
at large separations. Any scale dependence observed at the large scales
of the LSS correlation function would disprove all the simple single
slow-roll inflationary models \citep{creminelli_2010}, since they
predict a very small amplitude of non-Gaussianity. There are better
chances in detecting a divergence from Gaussianity by using  galaxy
clustering, since the signal is stronger than in CMB and matter
fluctuations can be examined at smaller scales. In addition, the galaxy
correlation function can probe the fluctuations at a time closer to the
present. 

Non-Gaussianities are expected in the LSS even if they are not present
in the primordial fluctuation, due to non-linearities in the structure
formation process. General relativistic correction can affect the
large-scales of the galaxies clustering, where \fnl constraints in LSS
surveys come from. \citet{maartens_2013}  showed that the
non-Gaussian signal dominates over gravity non-linearities for
$f_{\mathrm{NL}}^{\mathrm{local}}\gtrsim5$.

Recently LSS surveys have provided competitive constraints on the \fnl
magnitude relative to the one's coming from the CMB bispectrum
measurements. \citet{xia_nvss}  found
$25<f_{\mathrm{NL}}^{\mathrm{local}}<117$ at $95\%$ CL, where they
analysed the correlation function of extragalactic radio sources at
redshift $z\approx1$. \citet{slosar_2008}  measured
$-29<f_{\mathrm{NL}}^{\mathrm{local}}<70$ and \citet{xia_2011} 
found $5<f_{\mathrm{NL}}^{\mathrm{local}}<84$ , both at $95\%$ CL.
Finally less constrained results from LSS surveys can be found in
\citet{ross_aj2012,padmanabhan2007},
$-92<f_{\mathrm{NL}}^{\mathrm{local}}<398$ and
$-268<f_{\mathrm{NL}}^{\mathrm{local}}<164$ respectively and in
\citet{nikoloudakis2012}, $f_{\mathrm{NL}}^{\mathrm{local}}=90\pm30$,
all at $95\%$ confidence level except for \citet{nikoloudakis2012} where
the results are at $68\%\;CL$. 

In this work we will use the non-Gaussian bias to probe non-Gaussianity
by analysing the SDSS-III Data Release 9 \citep{SDSS_DR9} Baryon Oscillation
Spectroscopic Survey \citep{BOSS} quasar sample. The SDSS-III DR9 BOSS
consists of two spectroscopic surveys. The first survey will measure the
redshift of $1.5\times10^6$ colour-magnitude selected high-luminosity
galaxies up to $z=0.7$ with a magnitude limit at $i<19.9$. The second
spectroscopic samples of the SDSS-III BOSS, which we will use here, consists
of $150,000$ quasars, selected from roughly $400,000$ targets, in the
redshift range of $2.2 < z < 3.5$, with a median redshift of $z\sim2.5$.
Quasar clustering can shed light on critical matters of the galaxies
formation and evolution, as well as black hole growth, wind and feedback
models. Quasars can be excellent candidates for constraining primordial
non-Gaussianity, since they are high biased tracers and can be detected
at large redshifts due to their high luminosity. Such objects can
provide a better chance in finding any scale dependence in the
large-scales of their clustering, since the signature of an existing
extra non-Gaussian term in the bias would be more evident.

The outline of the paper is as follows. In Section 2 we describe the
SDSS DR9 BOSS quasar sample, quasar selection technique and angular
completeness. In Section 3 we present the clustering results, comparing
them with earlier works on the same quasar sample. In Section 4 we test
the quasar sample for any non-Gaussian signal putting constraints on
\fnl. In Section 5 we test and correct our sample for systematic errors
receiving new measurements on non-Gaussianity. We discuss our results
and conclude in Section 6.

Throughout this paper, we use a flat $\Lambda$-dominated cosmology with $\Omega_m=0.27$, $\Omega_{\Lambda}=0.73$, $H_{0}=100h\;kms^{-1}\mathrm{Mpc}^{-1}$, $h=0.7$, $\sigma_8=0.8$, $n_s=0.96$, unless otherwise stated.

\section{DATA}

In this work we use the data from  the Data Release Nine \citep{SDSS_DR9} of the SDSS-III BOSS survey. More precisely we use the dataset that applies the \emph{extreme deconvolution} algorithm (XD) of \citet{bovy2009,bovy2011} for the quasar target selection in order to identify objects for spectroscopic observation. After applying the XD method, every point
 source of SDSS-III BOSS is assigned with a XDQSO probability of being a
 quasar, by modelling the flux distribution of quasars and stars. In this
 way, a separation between targeted quasars and possible star
 contaminants is achieved. The details on the quasar selection is described in \citet{boss_qso_selec}. The complete catalogue of the DR9 SDSS-III BOSS spectroscopically confirmed quasars is presented in \citet{dr9_boss_qso}, where $61,931$ objects are included with redshift $z>2.15$.

 \subsection{Imaging and Quasar selection}
 
 The SDSS-III BOSS survey uses the data gathered using a dedicated 2.5 m wide-field telescope \citep{gunn_2006} to collect light for a camera with 30 $2k\times2k$ CCDs \citep{gunn_1998} over five broad bands, $ugriz$ \citep{fukugita_1996}. The camera has imaged $14,555$ unique $\rm{deg}^2$ of the sky, where $\sim7,500\;\rm{deg}^2$ are included in the North Galactic Cap and $\sim3,100\;\rm{deg}^2$ in the South Galactic Cap \citep{aihara_2011}. The imaging data were taken on dark photometric nights of good seeing \citep{hogg_2001}. After measuring the properties and parameters of the objects \citep{lupton_2001,stoughton_2002} the data are photometrically \citep{smith_2002,ivezic_2004,tucker_2006,padmanabhan_2008} and astrometrically \citep{pier_2003} calibrated.

 The redshift range of the quasar selection in the BOSS survey was
 selected to be $2.2<z<3.5$, since this is the sensitive region of the BOSS
 spectrograph for measuring Ly-$\alpha$ forest \citep{boss_qso_selec} as well as the number density of quasars is highly reduced at $z>3$ \citep{schmidt_1995,richards2006,ross_nic2013,mcgreer2013}. In addition to that at redshift $2-3$ a peak in
 the number density of luminous quasars has been observed
 \citep{richards2006,croom2009,ross_nic2013}. However, the quasar selection is
 complicated at these redshifts. At redshift $z=2.7$ the quasar
 colours are similar to the colours of metal-poor A and F star populations
 \citep{fan1999,richards2001} making the separation between these two even more
 difficult. Moreover quasars at redshift $z\sim2.5$ are contaminated
 from lower redshift ($z\sim0.8$) less luminous quasars that have
 similar colour and luminosity with them \citep{richards2001,croom2009}.
    
 In order to use the SDSS-III BOSS quasars for statistical analysis (i.e.
 clustering studies), we need to produce a uniformly selected sample. All the point sources of BOSS
 with XDQSO probability above 0.424 create the uniformly selected CORE sample
 \citep{boss_qso_selec}, which consists out of $74,607$ objects. In addition to this, a BONUS sample is also
 constructed by using as many additional data and techniques needed to
 reach the desired quasar density. More details on the CORE+BONUS method
 and the XDQSO technique used for the BOSS quasar selection,  together
 with the details of the pipeline used are listed by
 \citet{boss_qso_selec,dr9_boss_qso,white2012}. 
 
 These XDQSO CORE quasar targets are matched with the list of objects included in DR9 (i.e. $spAll-v5\_4\_45$) that BOSS successfully obtained a spectrum. In this way we create a final sample out of the matched objects ($63,205$), where each quasar target has a spectroscopically assigned redshift and classification together with additional photometric and other spectroscopic details. The same sample has been used before to analyse the clustering of quasars with $z>2.2$ in
    \citet{white2012}.

 \subsection{Sub-sample and angular completeness}

    \begin{figure}
    \centering
    \resizebox{\hsize}{!}{\includegraphics{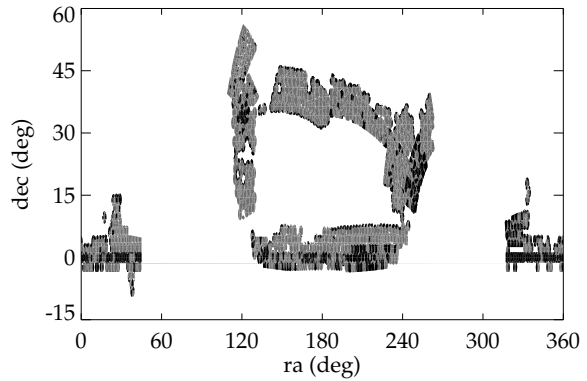}}
    \caption{The MANGLE mask after removing the veto mask for bright stars, bad u-band fields, bad photometry and spectroscopic plate's centerposts. The polygons observed contain the XDQSO CORE objects (mask is one). The grey polygons are those that meet the $75\%$ completeness threshold applied here, where the rest (black) do not and the objects inside them are removed from the sample.}
    \label{fig:qso}
    \end{figure}

    The MANGLE software \citep{mangle} is used to apply the angular mask of the BOSS DR9 survey. The ``one'' mask of our sample consists only from the MANGLE polygons that contain all the XDQSO CORE targets. A weight defining the completeness value inside each one is calculated. The angular completeness ($f_{\rm{comp}}$) is determined by the percentage of the targeting quasars, in a sector, that get a BOSS fibre to measure their spectrum. In this work, following \citet{white2012}, a 75 per cent threshold is set to the completeness. This simply means that we will keep the regions inside which $75\%$ and more of the targeted XDQSO CORE quasars have been assigned a fibre. We vary the value of $f_{\rm{comp}}$ to test the completeness threshold for being a potential systematic error, affecting the clustering of the quasar sample. Our finding together with a detailed analysis is presented in Section 5. In addition we remove regions with bright stars, since no quasars can be observed there, as well as regions with 
bad u-band data, bad photometry and spectroscopic plate's centerposts. After the application of the $f_{\rm{comp}}$ cut-off we have a sample of $51,584$ objects. More details on the above process, as well as on the redshift assignation of quasars from their spectrum and the redshift errors can be found in \citet{dr9_boss_qso,white2012}.
   
\begin{figure}
    \centering
    \resizebox{\hsize}{!}{\includegraphics{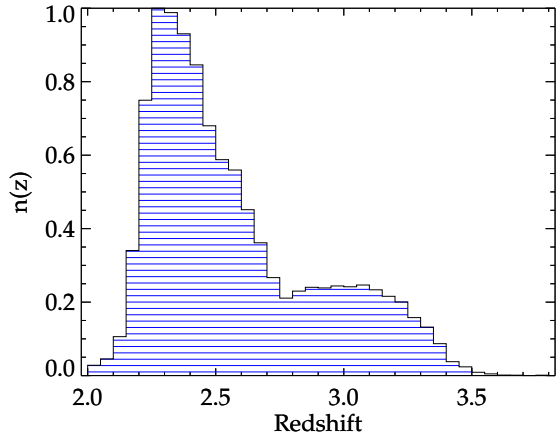}}
    \caption{The normalized redshift distribution, $n(z)$, of the BOSS CORE quasar sample.}
    \label{fig:qsonz}
    \end{figure}

   The resulting angular mask of the SDSS-III DR9 BOSS quasars is plotted in Fig. \ref{fig:qso}, where the regions that meet the $75\%$ completeness threshold are plotted in grey and the rest are plotted in black. We keep the objects in the redshift range of $2<z<3.8$ (Fig. \ref{fig:qsonz}). We end up with a sample of $29,687$ quasars, after keeping only the objects with $zWARNING=0$ indicating quasars with no known problem in their spectra. If the zWARNING flag \citep{adelman-mccarthy2008}, that is determined from the spectroscopic pipeline, is equal to zero then the redshift is accurate at $99.7\%$ level. Finally we observe in Fig. \ref{fig:qsonz} that most of the objects are concentrated in the redshift range $2.2<z<2.9$, with the peak being at $z\sim2.3$. Therefore, we will make a redshift cut and use only the quasars in redshift range $2.2<z<2.9$ for our analysis, leaving $22,361$ objects in the sample. This final quasar sample is the one that we will use here for our analysis. The properties of the 
sample used here are presented in Table \ref{table:sample_descr}. 

\section{CLUSTERING ANALYSIS}

   Galaxies are not randomly distributed in the universe, but rather form clusters and super-clusters through the attractive force of gravity. The probability of finding one galaxy at a point inside the volume $dV_1$ and an other one inside volume $dV_2$ , within a separation distance $r$ between them is
   
   \begin{equation}
    \delta P=n_V^2\left(1+\xi(r)\right)dV_1dV_2
   \end{equation}
   
   \noindent where $n_V$ is the mean number of galaxies inside volume $dV$ and $\xi(r)$ is the two-point correlation function. It provides a complete statistical characterization of the density field, as far as Gaussian statistics can be applied, by giving us information on the excess probability of galaxies being separated at different scales (clustering) compared to a randomly unclustered uniformly distributed sample. 
   
   There are many estimators calculating the two-point correlation function. The most common estimator used in literature, which is the one that we will also use here, is the Landy $\&$ Szalay estimator \citep{landy}
   
   \begin{equation} \label{eq:L&S}
   \xi(s)=1+\frac{DD(s)}{RR(s)}\left(\frac{N_R}{N_D}\right)^2-2\frac{DR(s)}{RR(s)}\left(\frac{N_R}{N_D}\right)
   \end{equation}

   \noindent where $DD(s)$ is the number of quasar-quasar pairs, $DR(s)$ is the pair counts of quasars and random point from the random catalogue and $RR(s)$ is the number of random-random pairs. All these pairs are counted in a bin of redshift space separation $s$ and over the entire survey area. The parameters $N_{rand}$ and $N_{dat}$ are the total number of random points and quasars respectively, and they are used as a normalization factor. The random catalogue has the same sky coverage as our data, as well as a smooth redshift distribution. Also it has to be large enough in order to reduce the Poisson errors. Therefore the random catalogues created here are $\sim20$ times bigger than the quasar sample, giving a normalization ratio $N_{rad}/N_{dat}\sim20$. Moreover a random catalogue $\sim50$ times bigger than the real data is constructed in order to test any difference between the two results. No statistical difference is observed between the two correlation functions and hence to save time in the 
calculation process we use the $\sim20$ times larger random catalogue.

     \begin{table}
    \centering
    \begin{tabular}{l|c}
    \hhline{==}
    Description & Number of Objects \\ \hline
    XDQSO QSO targets & $74,607$ \\
       " with spectra & $63,205$ \\
       " and $f_{comp}\geq0.75$ & $51,584$ \\
       " and $2<z<3.8$ & $30,681$ \\
       " and $\rm{zWARNING}=0$ & $29,687$ \\
       " and redshift range $2.2<z<2.9$ & $\mathbf{22,361}$ \\
    \hhline{==}
    
    \end{tabular}
    \caption{Properties of the SDSS-III BOSS XDQSO CORE sample.}
    \label{table:sample_descr}
    \end{table}
   
   The random points must be created in the regions where the completeness is above $75\%$, inside which the quasars of our sample are located. In order to achieve that we use the \emph{ransack} program of MANGLE to randomly generate angular coordinates of points inside the mask of the survey. To assign a redshift to each one of the random points we randomly take redshift values from the range of the quasar redshift distribution. As also noted by \citet{white2012}, this method can produce artificial structures in redshift distribution of the random points, since it follows the distribution of the data. However, with the large angular size of the BOSS survey this method gives correct results.
   
   \subsection{Error estimator}
    
    In order to determine the statistical uncertainty of the measured quasar correlation function, we will use the jackknife re-sampling method, which is an internal method of error estimation. The sample is split into $N_{sub}=159$ angular regions (subfields) of roughly equal size ($21.1\pm2.2\;\mathrm{deg}^2$) after taking into consideration the completeness mask (Fig. \ref{fig:qso}) as well as the area that each polygon covers. We then reconstruct copies of the data by omitting in turn one subfield at a time, hence creating $N_{sub}$ different realizations of the original sample. The main idea of the jackknife re-sampling is to measure the correlation function of each realization and compare it with the mean correlation function of all the realizations, which in fact is the correlation function of the original data set. The jackknife error estimator is given by
    
    \begin{equation} \label{eq:jk}
    \sigma_{jk}^2(s)=\frac{N_{sub}-1}{N_{sub}}\sum_{i=1}^{N_{sub}}\left[\xi_i(s)-\xi(s)\right]^2
    \end{equation}
    
    \noindent where the factor, $(N_{sub}-1)/N_{sub}=158/159$, takes into account the fact that the different realizations are not independent \citep{ross_percival_2011,crocce2011}. The sum is over the square of the difference between the sample's correlation function measured without the $ith$ subsample ($ith$ realization) and the correlation function measured from the whole quasar sample. The jackknife error technique has been used before in many clustering analysis studies, such as \citet{zehavi2005a,ross_n2007,sawangwit2011,nikoloudakis2012}. A detailed analysis on the error estimators for two-point correlation functions can be found in \citet{norberg2008}.
    
    \begin{figure}
    \centering
    \resizebox{\hsize}{!}{\includegraphics{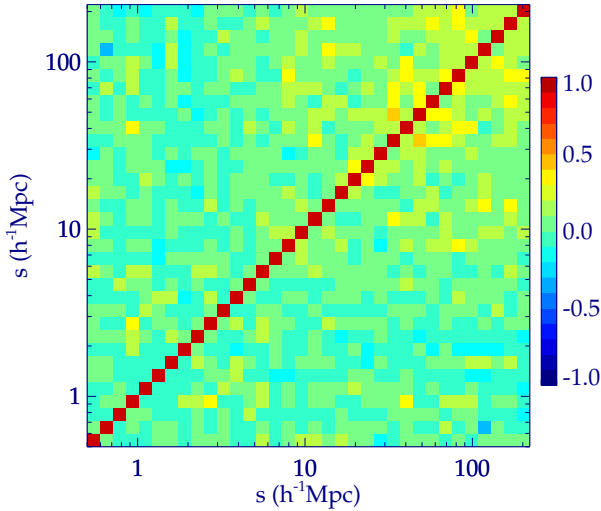}}
    \caption{The correlation coefficient $\mathbf{r}_{ij}$, which shows the level of correlation between each bin of separation s, where $+1$ means that they are fully correlated and $-1$ anti-correlated.}
    \label{fig:coef}
    \end{figure}     
    
    The main purpose of this work is to constrain non-Gaussianity from the BOSS quasar sample. To do this we will fit generated models that incorporate the non-Gaussian bias to the observed correlation function of the quasars. Hence in order to get accurate and robust results from the fitting we have to calculate the full covariance matrix from
    
    \begin{equation} \label{eq:covar}
     \mathbf{C}_{ij}^{jk}=\frac{N_{sub}-1}{N_{sub}}\sum_{k=1}^{N_{sub}}\left[\xi_i^k-\overline{\xi_i}\right]\left[\xi_j^k-\overline{\xi_j}\right]
    \end{equation}
    
    \noindent where $\overline{\xi}(s)$ is the mean correlation function of all the realizations, $\xi_i^k(s)$ is the correlation function of the sample without the $kth$ subsample and the subscript is the bin number. It is easy to understand that the jackknife error estimator (Eq. \ref{eq:jk}) is just the diagonal elements of the covariance matrix, $\sigma_i^{jk}=\sqrt{\mathbf{C}_{ii}^{jk}}$. We can now compute the \emph{correlation coefficient}, $\mathbf{r}_{ij}$, as defined from

    \begin{equation}
     \mathbf{r}_{ij}=\frac{\mathbf{C}_{ij}}{\sqrt{\mathbf{C}_{ii}\cdot\mathbf{C}_{jj}}}
    \end{equation}

    \noindent which is plotted in Fig \ref{fig:coef}. As we can see, the correlation of the different separation bins is negligible at small scales, while at larger scales it is higher but still not significantly large. The matrix is diagonal-dominated as expected.

    \subsection{Clustering results}
   
\begin{figure}
   \centering
   \resizebox{\hsize}{!}{\includegraphics{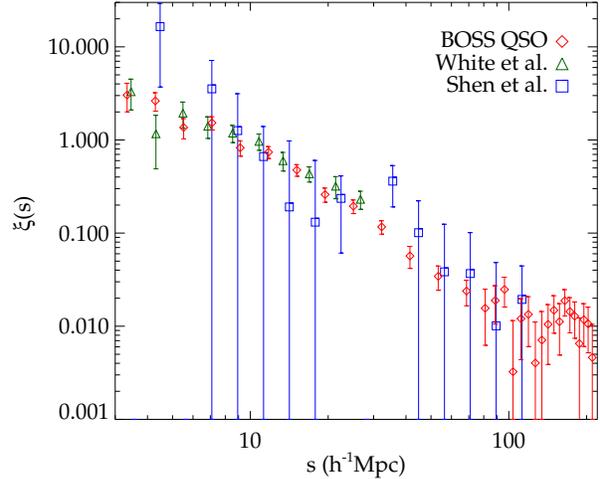}}
   \caption{The clustering results of the BOSS quasar sample, where the errors are the square root of the diagonal elements of the jackknife covariance matrix, together with the clustering results of \citet{white2012}, which uses the same quasar sample but with different redshift cuts ($2.2<z<2.8$), and with the redshift-space correlation function of the SDSS DR5 quasars from \citet{shen2007}, within the redshift range of $2.9<z<5.4$.}
   \label{fig:clust}
   \end{figure}

    The two-point correlation function of the quasar BOSS sample is measured in redshift space, by using the estimator of Eq. \ref{eq:L&S}. To count the pairs needed in the Landy $\&$ Szalay formula we use the kd-tree code of \citet{moore2001_npt}. The 3-D correlation results of the quasars are plotted in Fig. \ref{fig:clust}. In order to compare our results with the clustering results of other high redshift quasar samples we plot in the same figure the correlation function of the same BOSS CORE quasars as measured in \citet{white2012}, with redshift $2.2<z<2.8$, as well as the results from the quasar sample analysed in \citet{shen2007}, within the redshift range of $2.9<z<5.4$.
    
    The sample of \citet{white2012} is the same XDQSO CORE quasar sample we use, with the same selection techniques and redshift distribution. The quasar sample analysed in \citet{shen2007} consists of 4,426 luminous optical quasars from SDSS DR5 at redshift range $2.9<z<5.4$. The error bars in these two samples is the jackknife re-sampling, which is the same error estimator as the one applied in our sample.

    The redshift space correlation function of BOSS quasars measured in \citet{white2012} is measured only for the small scales, $3<s<30\;\mathrm{h}^{-1}\mathrm{Mpc}$. As we can see in Fig. \ref{fig:clust}, it is in very good agreement within the uncertainty limits, implied by the covariance matrix, with the correlation function measured here by using the same quasar sample. The slight differences can be attributed to the different redshift cut applied in the sample of \citeauthor{white2012}. The \citeauthor{shen2007} quasars are one of the highest redshift clustering studies, with all the quasars having $z>2.9$. Most of the correlation function points in \citeauthor{shen2007} reside near the correlation function measured from the BOSS quasar sample, the huge error bars of the first doesn't allow us to make a fair comparison between the two. The two redshift distributions do not overlap, where our quasar sample has redshift within $2.2<z<2.9$ while in \citeauthor{shen2007} the redshift range is $2.9<z<5.4$. 
However both redshift space correlation functions are consistent with each other within the uncertainty boundaries of their jackknife errors (see Fig. \ref{fig:clust}). We can conclude that the measured correlation function of the BOSS quasar sample, measured here is consistent with the results of \citeauthor{white2012} and \citeauthor{shen2007}, where they analyse the clustering of quasar samples in the redshift range similar or close to the one we are using in this work.

    At this point we have to note that the selection of the $\delta s$ bins has been such that the first 20 bins have an increasing step which becomes fixed at large separations ($s>80\;\mathrm{h}^{-1}\mathrm{Mpc}$). This binning has been chosen in order to acquire a more detailed description of the large scales of clustering, which are at most importance for our analysis on non-Gaussianity. Different binning has been also tested (e.g. steady step value for all the scale range, increasing binning step for the small scales and a steady but smaller one for the large, binning of the large scales in one and two points) but it does not affect the results of our analysis.

    \subsection{Modelling and fitting}
    
    Ordinary matter trapped in the gravitational well of the dark matter halos will cool and concentrate to create galaxies. This galaxy formation process introduces a bias relation between the galaxy and the underlying dark matter distribution \citep{kaiser_1984,fry}

    \begin{equation} \label{eq:bias}
    \xi_g(r)=b_E^2\xi_m(r)
    \end{equation}

    \noindent where $\xi_g(r)$ and $\xi_m(r)$ is the two-point correlation function at scale $r$ of the galaxy and the dark matter distribution respectively. The Eulerian galaxy bias is related to the Lagrangian bias via, $b_E=1+b_L$. To measure the linear bias of the sample (Eq. \ref{eq:bias}) we have to fit a generated model that follows the standard $\Lambda$CDM cosmology to the clustering results of the quasar sample.
    
    The $\Lambda$CDM model is created by generating an initial linear matter power spectrum, which as being the Fourier transformation of the two-point correlation function, will give $\xi_m(r)$ (Eq. \ref{eq:bias}). We use the formulas described in \citep{eh1998} to construct the matter power spectrum

    \begin{equation} \label{eq:model}
    P_m(k)=Ak^{n_s}T^2(k) 
    \end{equation}
    
    \noindent where $n_s$ is the spectral index of the usual power-law power spectrum. The transfer function $T(k)$ used is the one defined in \citep{eh1998}. The normalization constant $A$ normalizes the power spectrum at $z = 0$ to give \mbox{$\sigma_8 = \sigma(R=8\;\mathrm{h}^{-1}\mathrm{Mpc}) = 0.8$}, with $\sigma(R)$ being the smoothed variance of the initial density field at scale $R$. It is given by
    
    \begin{equation}
     A=\frac{1}{2\pi^2}\frac{\sigma_8}{\int_0^{\infty}P_m(k)k^2W(k\cdot8\;\mathrm{h}^{-1}\mathrm{Mpc})}
    \end{equation}
    
    \noindent where $W(kR)=3\left(\sin(kR)/kR+\cos(kR)\right)/(kR)^2$, is the Fourier coefficient of the spherical top-hat window function. From the Fourier transformation of the matter power spectrum we acquire the matter two-point correlation function
    
    \begin{equation} \label{eq:cf_model}
    \xi_m(r)=\frac{1}{2\pi^2}\int_{0}^{\infty} P_m(k)\frac{\sin(kr)}{kr}k^{2}dk
    \end{equation}
    
    \noindent To linearly extrapolate the matter correlation function to $z=0$ we follow the linear theory, where the evolution of structures is described by the growth factor, $D(z)$. Therefore, $\delta(r,z)=D(z)\delta(r,z=0)$, with $\delta(r)$ being the overdensity of the matter field. From the definition of the correlation function, $\langle\delta(\mathbf{x})\delta(\mathbf{y})\rangle=\xi(|\mathbf{x}-\mathbf{y}|)=\xi(r)$, we have finally
    
    \begin{equation}
     \xi_m(r,z)=D^2(z)\xi_m(r,z=0)
    \end{equation}
    
    \noindent It is obvious that the growth factor in the present time is unity.
    
    To generate the $\Lambda$CDM modelled correlation function and measure the best-fit bias, we need to define an average redshift which we will use for our calculations. According to \citet{white2012} if we cut our sample in redshift bins, big enough for bias to change from the one bin to the other, and calculate the correlation function in each bin we will define a redshift averaged $\xi(r)$. This correlation function is equivalent with the $\xi(r)$ calculated at an effective redshift, $z_{eff}$, defined as
    
    \begin{equation}
     z_{eff}=\frac{\int dz\; n^2(z) (H(z)/d_A^2)z}{\int dz\; n^2(z) (H(z)/d_A^2)}
    \end{equation}

    \noindent where $n(z)$ is the redshift distribution, as shown in Fig. \ref{fig:qsonz}, $d_A$ is the comoving angular diameter distance and $H(z)$ is the Hubble parameter at redshift z. The effective redshift of our sample is $z_{eff}=2.4$ and is the redshift we will use in our calculations and it is consistent with the redshift measured in \cite{white2012}, which is $z_{eff}=2.39$.
    
    Finally in order to measure the linear bias we have to take into consideration that the generated $\Lambda$CDM model gives a correlation function in the real space, while the measured one is in the redshift space. Hence the relation we will use to calculate the best-fit bias, that incorporates the analogy between the two distributions, is according to \citet{kaiser1987}
    
    \begin{equation} \label{eq:kaiser}
     \xi_{\mathrm{qso}}(s)=\left(b^2+\frac{2}{3}bf+\frac{f^2}{5}\right)\xi_m(r)
    \end{equation}
    
    \noindent where $f=\Omega_m(z)^{0.56}$, is the gravitational growth factor. 
    
    The model is fitted to the measured correlation function $\xi_{\mathrm{qso}}(s)$ by minimizing the $\chi^2$ statistics with the full covariance matrix, calculated from
    
    \begin{equation} \label{eq:chi}
    \chi^2=\sum_{i,j=1}^N(\xi_i-\xi_i^m)\mathbf{C}_{ij}^{-1}(\xi_j-\xi_j^m)
    \end{equation}
    
    \noindent where the sum is over the different bins $i$ and $j$, $\mathbf{C}^{-1}$ is the inverse of the covariance matrix defined from the jackknife re-sampling method (see Eq. \ref{eq:covar}), $\xi_i^m$ and $\xi_i$ is the value of the modelled and measured correlation function respectively, at the $i$th bin.

    The best-fit linear bias is estimated as being the only freeparameter of the generated model. We fit $\Lambda$CDM in the scales $3<s<50\;\mathrm{h}^{-1}\mathrm{Mpc}$, after taking into account the Kaiser effect. The resulting linear bias is found to be $b=3.74\pm0.12$ with $\chi^2_{red}=1.28$. While after fitting the model to the whole range of scales,
    $3<s<210\;\mathrm{h}^{-1}\mathrm{Mpc}$, we measure a bias parameter of $b=3.7\pm0.11$ with $\chi^2_{red}=1.78$. The difference between the two best-fit linear biases is negligible and inside their $1\sigma$ error limits. The measured linear bias is in good agreement with $b=3.8\pm0.3$ found by \citet{white2012}, after analysing the same quasar sample, as well as the bias results from other quasar clustering studies \citep{croom2005} at overlapping redshift ranges.

    The observed excess in large scales clustering led to a higher $\chi_{red}^2$ value of the best-fit bias from the whole range of scales. This is due to the poor fit of the model at these separations. As a result of the statistical uncertainties of the large scales data, the measured value does not have a significant difference compared to the measurement coming from the fitting on the small scales. In our calculations we will use the one originating from the fitting of the model till the scales of $s\sim50\;\mathrm{h}^{-1}\mathrm{Mpc}$, since we are interested in measuring the Gaussian part of the bias and at this range the $\Lambda$CDM Universe based model fits well.
    
   \begin{figure}
   \centering
   \resizebox{\hsize}{!}{\includegraphics{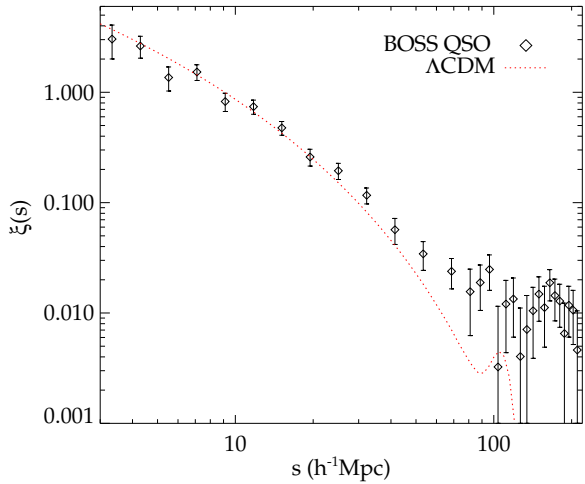}}
   \caption{The measured redshift space two-point correlation function, $\xi(s)$, for the BOSS quasar sample. The dotted line is the best-fit $\Lambda$CDM model, as defined in the text below.}
   \label{fig:clustmod}
   \end{figure}

    The resulting $\Lambda$CDM model together with the quasar correlation function is plotted in Fig. \ref{fig:clustmod}. As we can see the model fits well to the data  well up to  scales of $\sim50\;\mathrm{h}^{-1}\mathrm{Mpc}$. At scales larger than $100\;\mathrm{h}^{-1}\mathrm{Mpc}$ a plateau is observed that increasingly dominates  the quasar clustering. In addition a peak can be seen at $97\;\mathrm{h}^{-1}\mathrm{Mpc}$, rather than where it is expected at $\sim105\;\mathrm{h}^{-1}\mathrm{Mpc}$. However, it is detected at just the $1\sigma$ level. 
    
    We test the goodness-of-fit of the $\Lambda$CDM model to our data by fitting it in the scale range of $3-120\;\mathrm{h}^{-1}\mathrm{Mpc}$, where we measure $\chi^2_{red}=1.77$ for 18 degrees of freedom. The data in these scales reject the standard model at $2.2\sigma$ significance level. If we also include the large scales (i.e. $3-210\;\mathrm{h}^{-1}\mathrm{Mpc}$) in the fitting process the result of the goodness-of-fit will be, $\chi^2_{red}=1.78$ on 30 degrees of freedom. $\Lambda$CDM in this case is rejected at a significance level of $2.7\sigma$.

\section{TEST FOR NON-GAUSSIANITY}

    The presence of primordial non-Gaussianity affects the primordial gravitational potential perturbations (Eq. \ref{eq:phi_ng}), which will induce  non-Gaussian characteristics into the density field through the Poisson equation. This will affect the peaks of the initial matter density distribution, where the dark matter halos collapse on such overdensity peaks above a threshold $\delta_c$. Non-Gaussianities affect the high mass tail (rare events) of the halo mass function, where for a positive \fnl more high-sigma peaks will be generated leading to a larger number of high-mass dark matter halos. Besides the effect of non-Gaussianity in the mass function of halos, the existence of a gravitational potential bispectrum of the local type can introduce an extra scale dependent term in the dark matter halo bias and hence in the galaxy bias \citep{dalal_2008,MV08}, giving 
    
     \begin{equation} \label{eq:ngbias}
     b_{eff}^E(k,z,f_{\mathrm{NL}})=b_G^E+3f_{\mathrm{NL}}(b_G^E-1)\frac{H_0^2\Omega_m\delta_c(z)}{c^2T(k)k^2}
    \end{equation}
    
    \noindent where $b_G$ is the linear Gaussian bias measured in the previous section, $T(k)$ is the transfer function and \mbox{$\delta_c(z)=\delta_c^0(z)/D(z)\sim1.686/D(z)$} assuming an Einstein-de Sitte cosmology, with $D(z)$ being the linear growth factor normalized to be equal to unity at $z=0$. Here we follow \citep{NFW}, where they define the critical overdensity for the spherical collapse as $\delta_c(z)=\delta_c^0(\Omega_m(z))/D(z)$. For a $\Lambda$CDM cosmology it is $\delta_c^0=0.15(12\pi)^{2/3}\Omega_m^{0.0055}(z)$, with $\Omega_m(z)$ being the matter density at redshift $z$. Both formulae give roughly similar results at redshift $z_{eff}$.

    Moreover there is an additional scale independent term in the effective bias originating from the effect of primordial non-Gaussianity on the dark matter halo mass function \citep{giannantonio_2010,giannantonio_2012}. That additional term influences all scales but its amplitude is very small. Here we absorb that term in the Gaussian part $b_G^{E}$ of the effective bias. 
    
    The generated non-Gaussian model originates from Eq. \ref{eq:kaiser}, where after taking into consideration the Kaiser effect, we replace the linear bias with the effective non-Gaussian bias (Eq. \ref{eq:ngbias}). The matter model used is the same as in Section 3.3, with $\xi_m(r)$ coming from Eq. \ref{eq:model} and Eq. \ref{eq:cf_model} after linearly extrapolating at redshift $z_{eff}=2.4$. Since the Gaussian bias for the quasar sample was measured in Section 3.3, the only unknown parameter of our model is \fnl. 
    
   \begin{figure}
   \centering
   \resizebox{\hsize}{!}{\includegraphics{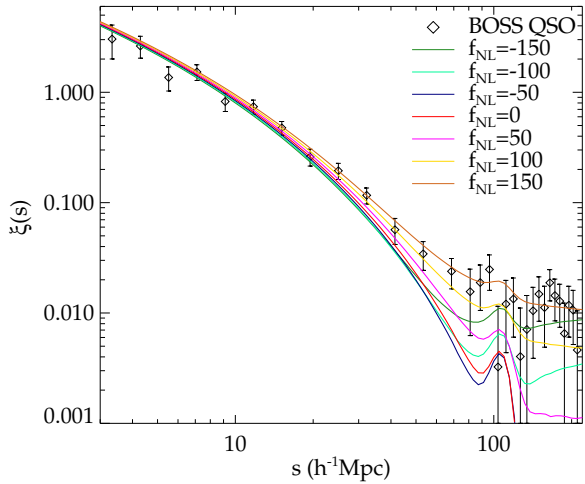}}
   \caption{The clustering results of the BOSS quasar sample. In red colour is the $\Lambda$CDM best-fit model ($f_{\mathrm{NL}}^{\mathrm{local}}=0$) after using the bias measured in Section 3.3. We have also plotted some non-Gaussian models, as described in Section 4, for different \fnl values. Both positive and negative \fnl value models fit well for the large scale plateau observed in the correlation function of the BOSS quasars. However the negative \fnl models are inconsistent with the clustering measurements between the scales of $20-90\;\mathrm{h}^{-1}\mathrm{Mpc}$.}
   \label{fig:nong}
   \end{figure}
    
    In Fig. \ref{fig:nong} we plot the generated model with the non-Gaussian bias for different \fnl covering a range from negative to positive values. In the same figure we plot for comparison the $\Lambda$CDM model, together with the measured quasar correlation function. All models have been calculated at the effective redshift, \mbox{$z_{eff}=2.4$}, and use the value of the best-fit linear Gaussian bias measured in the previous section ($b_{G}=3.74$). As analysed in the previous section the standard model (i.e. $f_{\mathrm{NL}}^{\mathrm{local}}=0$) fails to fit on the measured correlation function, since it goes too fast to zero while a flattening is observed in the clustering of the quasars beyond $110\;\mathrm{h}^{-1}\mathrm{Mpc}$. On the other hand non-Gaussian models, due to the scale dependent bias, can fit the observed large scale plateau, making them more consistent with our measurements (see Fig .\ref{fig:nong}). Both positive and negative \fnl value models fit well at the large scale excess, 
however the later ones are not consistent with the data in the scale range of $20-90\;\mathrm{h}^{-1}\mathrm{Mpc}$ as we can see in Fig. \ref{fig:nong}.   
    
    We use the full covariance matrix from the jackknife re-sampling, together with the minimization of the $\chi^2$ statistics (Eq. \ref{eq:chi}), in order to calculate the best-fit \fnl parameter and constrain the amount of primordial non-Gaussianity. The resulting value after fitting the model to the whole range of scales is, $f_{\mathrm{NL}}^{\mathrm{loc}}=131\pm17$ at $1\;\sigma$ CL with $\chi_{red}^2=1.15$.

    The number of free parameters in the model play a crucial role in the $\chi^2$ statistics and hence in the best-fit \fnl parameter and its variation. If we allow the Gaussian bias to be a free parameter, the $\chi^2$ test will give different \fnl with larger errors, since the smaller the Gaussian bias the larger the \fnl must be to fit our data and vice versa. This leads to bigger uncertainty of the \fnl value measured from this sample and hence weaker constraints. The same would have happened if we had calculated the quasar bias from the weighted average of the halo bias by considering a Halo Occupation Distribution (HOD) model. The plethora of different values of the free parameters of the best-fit HOD (e.g. the minimum mass of the halo) would give a bigger number of best-fitting combinations leading to higher uncertainty in the amplitude of non-Gaussianity. The fact that we measure the Gaussian bias from the best-fit of the $\Lambda$CDM model with the data leaves only one free parameter (\fnl) in the 
fitted non-Gaussian model. Therefore the measured \fnl is expected to be tightly constrained, since we consider all the other free parameters of the model well defined.

    In addition we have to take into consideration the fact that the measured Gaussian bias can vary inside the limits of its uncertainty affecting the value of $\chi^2_{min}$ and hence the uncertainty error of the best-fit \fnl. In order to calculate the new uncertainties of \fnl we allow the bias to vary inside its error limits, while at the same time we allow \fnl to vary around its best-fit value creating a grid. At every new set of $(b_G,f_{\mathrm{NL}}^{\mathrm{local}})$ we measure the new value of $\chi^2$, which will be larger than $\chi^2_{min}$ and corresponds to the best-fit set. We find the relation of $\Delta\chi^2=\chi^2-\chi^2_{min}$ with the variation of the two parameters $(\delta b_G,\delta f_{\mathrm{NL}}^{\mathrm{local}})$, which is nothing more than an ellipse (contour). In our case since we have put limits to the variation of one of the parameters the contour will not be complete. The maximum and minimum value of the $\Delta\chi^2=2.3$ contour\footnote{The $95\%$ confidence interval 
level for the two 
parameter fit is for $\Delta\chi^2=6.17$. The calculation of the uncertainty limits in the best-fit parameters from the confidence interval assume that the data errors follow a Gaussian distribution.} (since it follows a $\chi^2$ distribution) will give us the new uncertainty levels at $1\;\sigma$ for each parameter, where the limit values of $\delta b_G$ are the 
error limits of $b_G$ as measured in the previous section. The new uncertainty error for the best-fit \fnl parameter will be, $f_{\mathrm{NL}}^{\mathrm{loc}}=134\pm40$ at $1\;\sigma$ confidence level, which as expected is larger than the one measured for only one free parameter in the minimization of the $\chi^2$ statistics. 
 
    Comparing our results, $70<f_{\mathrm{NL}}^{\mathrm{local}}<190$ at $95\%$ confidence level, with the those measured in other galaxy samples \citep{slosar_2008} $-29<f_{\mathrm{NL}}^{\mathrm{local}}<70$, \citet{xia_2010} $10<f_{\mathrm{NL}}^{\mathrm{local}}<106$, \citet{xia_nvss} $25<f_{\mathrm{NL}}^{\mathrm{local}}<117$ and \citet{xia_2011} $5<f_{\mathrm{NL}}^{\mathrm{local}}<84$ all in $95\%$ CL, we find that the measured value of the non-Gaussian amplitude from the BOSS quasars has a small overlap with the values measured in the above studies. However this overlap is much more significant with the results of \citet{xia_2010,xia_nvss}. On the other hand our findings are consistent with the measurements coming from LSS clustering studies that give more loose constraints on the values of \fnl as in \citet{ross_aj2012,padmanabhan2007}, $-92<f_{\mathrm{NL}}^{\mathrm{local}}<398$ and $-268<f_{\mathrm{NL}}^{\mathrm{local}}<164$ respectively and in \citet{nikoloudakis2012}, \mbox{$f_{\mathrm{NL}}^{\mathrm{
local}}=90\pm30$} \footnote{\citet{nikoloudakis2012} suggest that their results should be considered as an upper limit on non-Gaussianity.}. All the previous results are at $95\%$ confidence level, besides the 
ones in \citet{nikoloudakis2012} where they are at $68\%\;CL$.

    The results of \citet{slosar_2008} come from the combined measurements of SDSS DR6 photometric quasars \citep{adelman-mccarthy2008} and LRG clustering, at redshift $1.5<z<2$ and $z\sim0.5$ respectively. \citet{xia_nvss} use the NVSS radio sources at redshift $z\sim1$ and \citet{xia_2010} use the cross-correlation of the NVSS with the SDSS DR6 QSO data. While \citet{xia_2011} use the combined results of NVSS radio sources, SDSS DR6 QSOs and MegaZ DR7 LRGs, with the first two having the same redshift ranges as those in the previous two studies and the MegaZ LRGs being in the redshift range of $0.4<z<0.7$. The SDSS DR6 quasar sample, that was used by \citet{xia_2010,xia_2011} to measure \fnl, was found to be very prominent to systematic effects \citep{pullen_2012} and that the sample is not fit for measurements on primordial non-Gaussianity. \citet{giannantonio_2013} argues, after further investigating the issue, that the quasars sample should be used only through cross-correlations with other surveys.
    
    These clustering studies that give \fnl constraints that just have a small overlap with the constraints measured here lie at lower redshifts, as well as using LRG and radio source data, objects that are less biased tracers than quasars. A large total bias can have an extra non-Gaussian term with a more significant contribution in the sum, giving a more distinctive non-Gaussian signature. The above facts could possibly explain the difference in the measured \fnl from the clustering of different tracers at smaller redshift ranges.  

\section{CHECK FOR SYSTEMATIC ERRORS}
    
    One of biggest disadvantage of constraining \fnl from LSS surveys is that they can be affected by potential systematic errors, which introduce fake signals and influence the clustering of the sample. Systematics can influence the whole range of the correlation function and mainly the highly sensitive region of large scales, where \fnl constraints originate. It is important to identify and correct any such sources, in order to acquire more robust constraints on non-Gaussianity. However this process turns out to be a difficult one.
    
   \begin{figure}
   \centering
   \resizebox{\hsize}{!}{\includegraphics{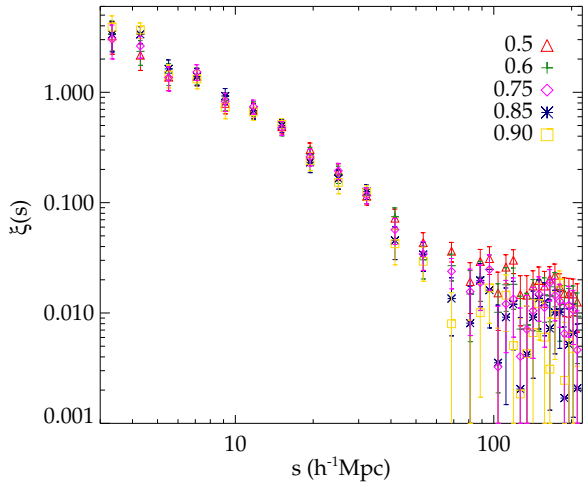}}
   \caption{The correlation function of the quasar sample after varying the value of the completeness threshold. Change in $f_{\rm{comp}}$ leads to a modification of the objects number included in the sample.}
   \label{fig:compl}
   \end{figure}
    
    We begin by changing the amount of completeness of our sample, as promised in Section 2.2, and measure each time the correlation function of the quasars in order to test how much does $f_{\rm{comp}}$ affects the clustering results. We note here that a change in the completeness will modify the number of the objects in the sample, since $f_{\rm{comp}}$ measures the percentage of quasars inside a sector of the BOSS mask that are assigned a fiber. The smaller this threshold is the more sectors are included in the sample leading to an increase in the objects number. More precisely after applying the same cuts as those applied for the main sample (see Section 2.2) we have $27,608$, $25,965$, $18,333$ and $15,826$ objects for $0.5$, $0.6$, $0.85$ and $0.9$ completeness cut-off respectively. We present our finding in Fig. \ref{fig:compl}, where we have plotted the correlation function of the sample after applying 5 different completeness threshold values. There is no statistically significant difference 
between the clustering results up to $\sim80\;\mathrm{h}^{-1}\mathrm{Mpc}$ by changing the completeness value. At the larger scales, where the clustering plateau is located, the divergence between the correlation functions is more evident but still insignificant and inside the $1\sigma$ uncertainty error. We conclude that completeness does not affect the clustering of our sample to a level where we can consider it a systematic error. Hence we stick to our initial choice for the completeness value, following \citet{white2012}.

    To test the quasar sample further for potential systematics we split it according to the value of different observational parameters. We divide the it into two parts by applying cuts on  galactic hemisphere, extinction, seeing and sky brightness. The correlation function of each part is measured, with the statistical errors being defined from the diagonal elements of the jackknife covariance matrix (see Section 3.1) as it is calculated from each individual part. The two resulting correlation functions from each different cut are compared in order to determine whether any of the applied cuts can affect the clustering of quasars and especially the large scales.
 
   \begin{figure}
   \centering
   \resizebox{\hsize}{!}{\includegraphics{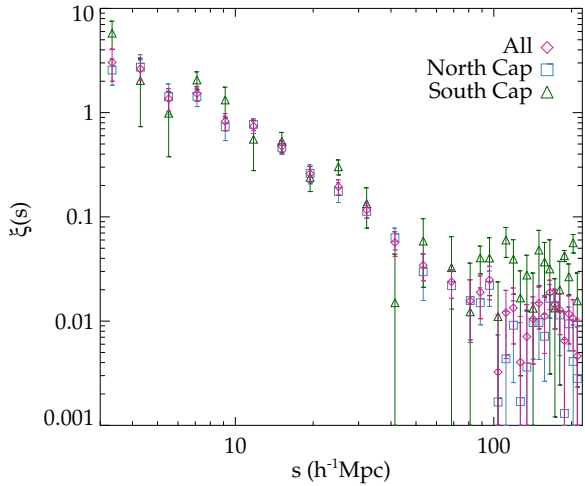}}
   \caption{The two point correlation function from the hemisphere-divided quasar sample is compared with the correlation function of the original sample. Note that the southern sample is much smaller than the northern sample. The  excess at large scales in the clustering of the South Cap sample  can be easily seen.}
   \label{fig:hemcut}
   \end{figure} 
 
    No statistically significant difference has been detected in any of the above cases, besides the north and south galactic hemisphere cut. In Fig. \ref{fig:hemcut} we present the correlation function of each part after dividing the sample into North and South galactic hemispheres, together with the original results from the full quasar sample. It is easy to see that the south galactic hemisphere quasars have a stronger clustering signal than those in the north galactic hemisphere, as well as the quasars from the full sample. This difference is located at  large  correlation function scales, especially at scales greater than $80\;\mathrm{h}^{-1}\mathrm{Mpc}$. But we note that the Southern Galactic quasar sample is roughly $5$ times smaller in area than the Northern sample and therefore more data are needed in the Southern hemisphere to check the reality of the observed difference between the clustering of the two samples.

\subsection{Quasar density vs. potential systematics}
    
    After the failure of these simple tests to point out any potential source of systematics, we will follow the simple error analysis of \citet{ross_2011,ross_2012,ho2012}, which is a widely used method in the literature for testing the robustness of the correlation results. In particular we will test if galactic extinction, seeing, sky brightness and stars can affect the large scales in such a way that they could falsely produce (at least partly) the observed excess in clustering, leading to a larger measured value of non-Gaussianity.
    
    \emph{Galactic extinction} caused by the presence of dust in the galactic plane, must be corrected in the measured magnitudes. Even though SDSS avoids most of the heavily extinct areas, errors in the correction precess can cause systematic errors \citep{scranton_2002,myers2006,ho_2008} affecting the galaxy density field and hence the clustering of the sample.
    
    \emph{Atmospheric seeing} is defined here as FWHM of the point spread function (PSF) and it is measured in arcseconds. It affects the de-blending of sources as well as the separation between extended and point sources (i.e. galaxy-star separation). 
    
    \emph{Sky brightness} measures the brightness of the sky during the time of the observation. It can affect the measured number of the objects of interest as well as their identification. It is measured in $nanomaggies/arcsec^2$ as taken from SDSS CAS.
    
   \begin{figure*}
   \centering
   \resizebox{\hsize}{!}{\includegraphics[trim=2cm 4cm 1cm 4cm]{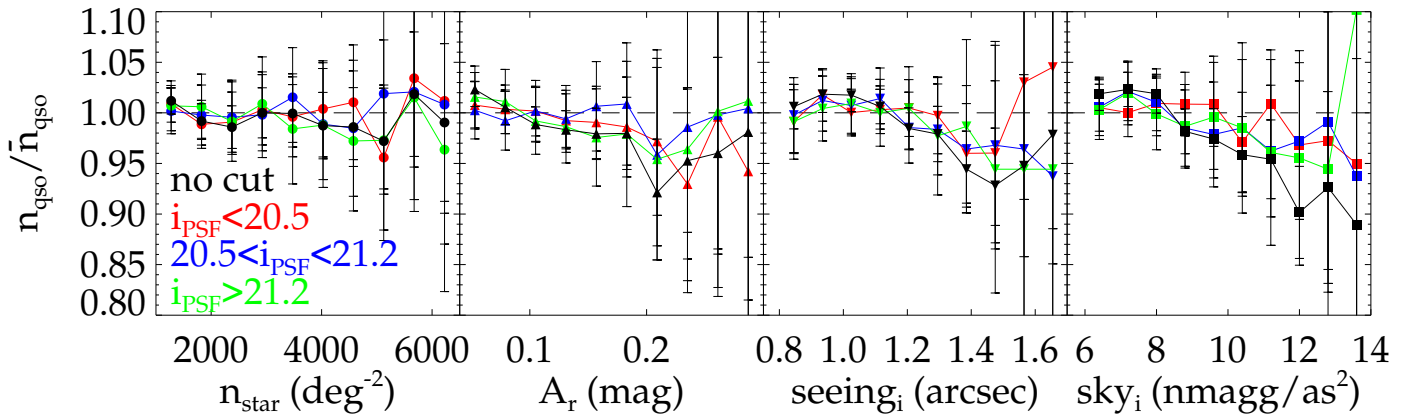}}
   \caption{The relationship between the number density $n_{\mathrm{qso}}$ of the BOSS DR9 quasar sample with respect to the average $\overline{n}_{\mathrm{qso}}$ and the potential systematics: stellar density ($n_{star}$), galactic extinction in the r-band ($A_{r}$), seeing in the i-band ($\rm{seeing}_i$) and the background sky flux in the i-band (sky$_i$). The black line displays number density of the whole quasar sample (no cuts applied) as a function of each potential systematic. The red, blue and green line represents the same relationship after the application of different cuts in the BOSS quasars in the PSF magnitude of the i-band, which are $i_{PSF}<20.5$, $20.5<i_{PSF}<21.2$ and $i_{PSF}>21.2$ respectively. The error bars in each point are Poisson.}
   \label{fig:histo_err}
   \end{figure*}
    
    \emph{Star contamination} can affect the number of objects measured in a survey. Stars can be falsely confused as quasars given they have the right colour, introducing a positive correlation between their density fields. In the case of photometric quasar surveys star contamination is even higher, since both objects are point sources and hence it is very difficult to identify them. Such stellar contamination effects together with the offset observed in the stellar colour locus have led \citep{pullen_2012,giannantonio_2013} to conclude that photometric quasar samples should be used through cross-correlations with other surveys in order to derive constraints on \fnl. The presence of systematic errors in such surveys can affect the  large scale correlation function  mimicking the signature of non-Gaussianity.

    \emph{Foreground stars} can also reduce the number of detected galaxies (or quasars in our  case) since there is a negative correlation between the number density of stars and that of galaxies as first observed by \citet{ross_2011}. It was found that the number density of galaxies drops $10\%$ from regions of high to regions of low stellar density. $3\%$ is caused, as stated by \citet{ross_2011,ross_2012}, by the fact that galaxies close to stars are not easily detectable. The other $7\%$ may come from the change in the photometric pipeline from DR7 to DR8. This change leads to the difficulty of the de-blending code to separate more than 25 overlapping objects in regions of high stellar density. Their analysis was for the SDSS-III BOSS DR9 CMASS sample of LRGs. Our data come also from the SDSS-III BOSS DR9 and hence such a systematic
    effect may be present also in the BOSS quasar sample. \citet{ross_2011,ross_2012} both find that the systematic effect
    from the foreground stars is the most important, and to correct it they finally use a method where they apply weights to the sample.

    In addition to the above possible systematics, it is important to consider the case of fibre collision. In the BOSS survey we cannot take the spectra of two quasars if their separation is smaller than $62''$, since the BOSS fibres cannot be placed closer than this distance. This leads us to miss some quasar-quasar pairs at these separations, affecting the small scales of the resulting clustering. The usual way to account for the colliding fibre issue is to weight up pairs of quasars with smaller separation than this, which in the case for $z\sim2.5$ is $1.26\;\mathrm{h}^{-1}\mathrm{Mpc}$.
    \citet{white2012} correct for such an effect following the weighting method \citep[see][]{ross_2011,ross_2012}, but they find no significant effect on the clustering. Our analysis focuses at the large scales of clustering which are not affected by the fibre collisions. In addition to this we only  include in our results scales greater than $3\;\mathrm{h}^{-1}\mathrm{Mpc}$ correcting in this way for the fibre collisions.
    
    We measure the relationship between the number of observed quasars and the potential systematics in order to find if any correlation exists between them. To achieve that, we pixelise the BOSS DR9 quasars, by using HEALPix \citep{gorski2005} with $N_{side}=256$, creating a map of our sample with equal area pixels of $0.0525\;deg^2$. In each pixel we calculate the mean of each potential systematic based on the SDSS DR9 CAS, as well as the number of quasars. In addition we measure the number of stars with $17.5<i_{mod}<19.9$ from the SDSS DR9 that reside in each of the above pixels after using the same mask as in the BOSS sample, since the stars that can affect our sample are those that overlap with it. Finally we measure the 
relationship of $n_{\mathrm{qso}}/\overline{n}_{\mathrm{qso}}$ with each systematic following the error analysis done by \citet{ross_2011,ross_2012,giannantonio_2013}, where $\overline{n}_{\mathrm{qso}}$ is the average number of quasars over all pixels.
    
    In Fig. \ref{fig:histo_err} we plot the number density of the quasar sample over the average number of quasars as a function of each potential systematic. In the first plot we can see (black line) the relationship between the quasar density and the density of stars. We do not observe any significant correlation between the two, besides a very small reduction in the number of quasars for $n_{star}>3500\;deg^{-2}$ equal to roughly $3\%$ followed by a rapid increase of $4\%$. At smaller values of $n_{star}$ the fluctuations are very small and around the best value
    of $n_{\mathrm{qso}}/\overline{n}_{\mathrm{qso}}(n_{star})=1$ without any particular trend. The observed small anti-correlation could be explained from the presence of the foreground stars as suggested by \citet{ross_2011}, where they find also a $3\%$ decrease in the number density of galaxies due to a masking of $10''$ from each star in the survey area. Inside that radius the seeing makes it less likely to detect any object reducing the number of observed quasars. The anti-correlation between the two samples has a magnitude dependence observed after applying different cuts on the
    i-band PSF magnitude of the quasars, where the faintest sample has a maximum reduction of $4\%$. As also explained by
    \citet{ross_2011} the less bright objects are more affected by the presence of foreground stars. 
 
   No positive correlation trend between the number of quasars and stars (see Fig. \ref{fig:histo_err}) is present, indicating a low percentage of stellar contamination. We only see a sudden increase of $10\%$ after $5100\;deg^{-2}$, which has more of a fluctuating nature. The highest increase in the quasar number density is observed for the brightest sample, but we can find the same trend after the application of the other magnitude cuts. As noted before, photometric quasar samples are highly affected by contamination, something though that doesn't apply in our case since the BOSS DR9 quasar sample originates from a spectroscopic survey. This means that the objects included in the sample are classified from their spectrum, which reduces stellar contamination to the minimum.
   
   In the other panels of Fig. \ref{fig:histo_err} we can see that the number density of quasars varies with the observational parameters of Galactic extinction, seeing and sky brightness. We find an anti-correlation between the Galactic extinction in the r-band and the number of quasars for $0.05-0.2 mag$ followed by a positive correlation. \citet{ross_2011} explain that such an anti-correlation is partially due to the fact that stars and extinction are correlated, since they both trace the structure of the Galaxy. Here we detect a $10\%$ decrease in the number density of quasars, which cannot be fully explained from the star-extinction correlation. The presence of the dust in the Galactic plane affects the number of detected quasars, where it appears here as a sharp decrease in the number density-extinction relationship from regions of small to regions of high Galactic extinction. After applying magnitude cuts to the quasar sample we find a non-trivial relationship between extinction and $i_{PSF}$, where 
the amplitude of the anti-correlation is reduced and the number density approaches its best-fit value ($n_{\mathrm{qso}}/\overline{n}_{\mathrm{qso}}(A_r)=1$) as we reach the magnitude limits of the survey.

   We detect a negative correlation between i-band seeing and number density of quasars for $0''.9$ and $1''.5$, followed by a slight increase of almost $5\%$ from regions of seeing $1''.5$ to regions of $1''.7$ (see Fig. \ref{fig:histo_err}). The explanation for the decrease lies in the fact that in regions of poor seeing the galaxy/star separation cut is affected. Such a cut is applied in the BOSS survey in order to separated extended from point like sources, where as explained by \citet{ross_2011} for an increasing seeing the PSF and model magnitudes become more alike. This happens because the
   PSF magnitude becomes more extended approaching the model one, therefore in poor seeing areas more objects that are point-like will be mistaken for extended sources by the applied cut. This would lead to an increase in the rejected objects leaving the sample with less point sources and hence potential quasars. Similar to the case of the
   systematic caused by Galactic extinction, we observe also here a relationship between the effect of seeing in the number of quasars and the i-band magnitude of the quasars.

   Finally we observe an anti-correlation between the number of quasars and the background sky flux in the i-band. We find a $13\%$ decrease in the quasar number density from regions with sky background of $7\;nmaggies/arcsec^2$ to regions with $14\;nmaggies/arcsec^2$. Again here we detect a non-trivial dependence of the ``number of quasars-sky flux'' relationship and the applied magnitude cuts, where the number density of the brighter
   sample seems to be less affected by the regions of high sky background values and almost approaching the best value of
   $n_{\mathrm{qso}}/\overline{n}_{\mathrm{qso}}(sky_i)=1$. High sky brightness regions can affect the identification of objects reducing the number of detected quasars, which can be identified in Fig. \ref{fig:histo_err} as a reduction in the number density with larger sky brightness values.

   \subsection{Extinction cuts}

   High extinction can introduce falsely artificial signals in the sensitive scales of the sample's correlation function. Here we find a decrease (almost $10\%$) in the number of quasars as the the value of Galactic extinction increases (see black line in Fig. \ref{fig:histo_err}). In the light of this observation we would like to test the effect of this systematic on the clustering result further, by applying stricter extinction cuts. Now instead of dividing the sample according to an extinction value (see Section 5), we only keep regions with $A_r<2.0$, $A_r<0.16$ and $A_r<0.14$ mag producing a sample of $20,800$, $18,521$ and $16,351$ quasars respectively. The removal of such regions can lead to the reduction of the large scale clustering and to a smaller measured \fnl value. 
   
   \begin{figure}
   \centering
   \resizebox{\hsize}{!}{\includegraphics{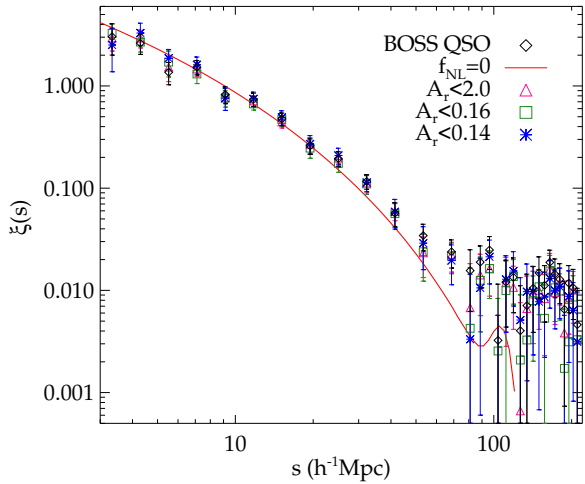}}
   \caption{The correlation function of the full sample together with
   the clustering from the reduced quasar sample after keeping
   regions with extinction $A_r<2.0$, $A_r<0.16$ and $A_r<0.14$ mag. The
   errors are calculated from the jackknife re-sampling method. We plot
   also the standard $\Lambda$CDM model
   ($f_{\mathrm{NL}}^{\mathrm{local}}=0$) with the linear bias measured
   in Section 3.3.}
   \label{fig:extcut}
   \end{figure}
   
   In Fig. \ref{fig:extcut} we plot the correlation function of the quasars after applying the three different extinction cuts together with the fiducial clustering results. The strict extinction cuts do not produce any significant change to the small and intermediate clustering scales. The large scales clustering is slightly reduced and more noisy than the fiducial results from the whole sample (see Fig. \ref{fig:clust}), but consistent inside the $1\sigma$ uncertainty limits. The highest significant difference from the fiducial clustering occurs after applying the strict cut of $A_r<0.14$, hence further on in this section we will use these results for our analysis.

   The $f_{\mathrm{NL}}^{\mathrm{local}}=0$ model is compatible with the data \mbox{till $\sim120\;\mathrm{h}^{-1}\mathrm{Mpc}$}, which is higher than the consistency limit of the standard model with the clustering results of the full quasar sample. A similar to the fiducial result's plateau is observed at large scales for all the extinction cuts with a small reduction to its amplitude, which cannot be fitted by the standard cosmological model. More precisely for scales $3-120\;\mathrm{h}^{-1}\mathrm{Mpc}$ the goodness-of-fit of the $f_{\mathrm{NL}}=0$ model is, $\chi^2_{red}=1.72$, which which rejects $\Lambda$CDM at the $2.1\sigma$ level. Including the large scales ($3-210\;\mathrm{h}^{-1}\mathrm{Mpc}$), the goodness-of-fit value is $\chi^2_{red}=1.39$. This indicates that data reject $\Lambda$CDM at the $1.7\sigma$ level. These goodness-of-fit results show that the standard model cannot be rejected at a high significance level from the data set after the application of 
the strict extinction cut.

     With this in mind  we fit to the whole scale range a non-Gaussian model to acquire new constraints on \fnl. After applying the minimization of the $\chi^2$ statistics method by using the full covariance, described in Section 3.3, we measure the best-fit non-Gaussian amplitude, \mbox{$f_{\mathrm{NL}}^{\mathrm{local}}=126\pm29$} at $68\%\;CL$ with \mbox{$\chi^2_{red}=0.9$}. The $95\%$ confidence level results are $77<f_{\mathrm{NL}}^{\mathrm{local}}<170$. The measured Gaussian bias is also used here, where we treat it as a free parameter allowing it to vary inside its error limits.  The new reduced \fnl may be lower than the one coming from the full sample, but still larger than zero. The results at the $95\%$ CL, overlap with the tight constraints on \fnl by \citet{xia_2010,xia_nvss,xia_2011}. Fig. \ref{fig:extcut} shows that the systematic error caused by Galactic extinction cannot individually generate falsely such a large excess giving an explanation for the observed clustering plateau, but it can be 
the cause for a part of it.

   \subsection{Correct the potential systematic sources}
   
   \subsubsection{Weights method}

      \begin{figure*}
   \centering
   \resizebox{\hsize}{!}{\includegraphics[trim=2cm 4cm 1cm 4cm]{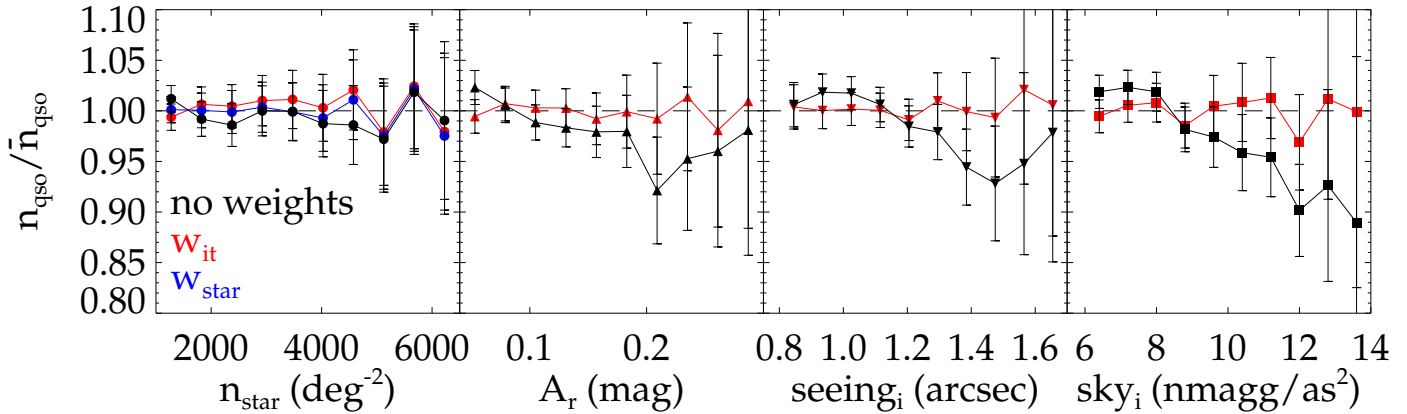}}
   \caption{The relationship between $n_{\mathrm{qso}}/\overline{n}_{\mathrm{qso}}$ and the potential systematics as in Fig. \ref{fig:histo_err}, where now we plot the function from the full sample (``no cut'' line in Fig. \ref{fig:histo_err}) and the corrected one after applying the iterative weights (red line). In the first panel we also plot the corrected relationship between number density and stars after we only apply weights that correct only the systematic caused by stars. The error bars in each point are also here Poisson.}
   \label{fig:histo_err_cor}
   \end{figure*}

    A sophisticated way to correct potential systematics that are
    present in a survey is by following the ``weight'' method developed
    by \citet{ross_2011,ross_2012}. The application of this approach is
    pretty straightforward, we apply weights to each quasar in order to
    remove any fluctuations in their number density with respect to each
    systematic (see Fig. \ref{fig:histo_err}). The weights are the
    reciprocal of the
    $n_{\mathrm{qso}}/\overline{n}_{\mathrm{qso}}\mathrm{sys}$
    relationship, plotted in black colour in Fig. \ref{fig:histo_err},
    where they are applied to the objects inside each Healpix pixel that
    belong to every systematic's bin. We start by applying the weights
    for the first systematic and then re-calculate the
    $n_{\mathrm{qso}}/\overline{n}_{\mathrm{qso}}\mathrm{sys}$
    relationship for the next systematic. The inverse of the function is
    then multiplied with the weights that the sample already has. This
    process is continued till we apply to the sample all the weights
    correcting for every potential systematic error considered here,
    creating what is referred in the Appendix of \citet{ross_2012} as
    iterative weights ($w_{it}$). To calculate $w_{it}$, the order for
    the applied weights is star, Galactic extinction, seeing and sky.
    The $n_{\mathrm{qso}}/\overline{n}_{\mathrm{qso}}\mathrm{sys}$
    relationship does not need to be linear in order to apply the
    weights method, which increases the simplicity of this technique.

    In Fig. \ref{fig:histo_err_cor} we plot the relationship of the quasar's number density as a function of the potential systematics before and after the application of $w_{it}$. The applied weights completely correct almost every fluctuations that appear in the $n_{\mathrm{qso}}/\overline{n}_{\mathrm{qso}}\mathrm{sys}$ relationship. This indicates that this systematic correction technique is too aggressive, since we expect more variations around unity \citep{ross_2012}.

    The weights method assumes that all the systematic sources are separable, which would be the case if the $n_{\mathrm{qso}}/\overline{n}_{\mathrm{qso}}\mathrm{sys}$ relationship was consistent with unity after applying $w_{it}$. In addition the order of the applied weights would not matter in the case of uncorrelated errors. However this assumption is not completely true. In the first panel of Fig. \ref{fig:histo_err_cor} we plot the corrected relationship with respect to the number density of stars after applying the $w_{it}$ and $w_{star}$ weights. The latest corrects the sample only for the systematic error caused by the presence of stars. The function with only $w_{star}$ applied is consistent with unity, where after the application of $w_{it}$ a small increase is observed. This means that we have a correlation between some of the systematic sources.

    After changing the application order of weights and repeat the
    process we find that the $w_{star}$ and $w_{A_r}$ are not completely
    separable. This is expected, as we explained before, since stars and
    Galactic extinction both follow the structure of the Galaxy
    \citep{ross_2011}. However since the fluctuations caused in the
    $n_{\mathrm{qso}}/\overline{n}_{\mathrm{qso}}(n_{\mathrm{star}})$
    relationship after the application of $w_{it}$ are small enough (see
    Fig. \ref{fig:histo_err_cor}), we can assume that these two
    systematics are separable. No other significant correlation is
    observed between the other systematic sources. 
    
\begin{figure}
   \centering
   \resizebox{\hsize}{!}{\includegraphics{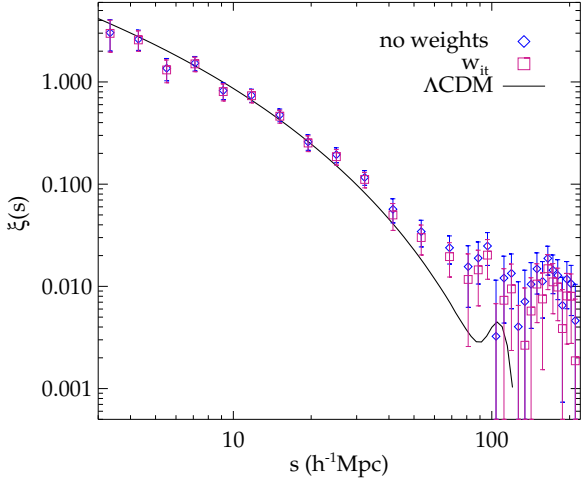}}
   \caption{The weighted correlation function after applying $w_{it}$ in each quasars of the sample. We also plot the fiducial (no weights) results together with a model following the standard cosmology.}
   \label{fig:clust_weight}
   \end{figure}

    The weighted correlation function of the data is calculated by following Eq. \ref{eq:L&S}, where now the number of pairs inside each separation bin is the sum of the product between the weights of the objects in each pair. The weighted clustering results are plotted together with the fiducial ones and the generated $\Lambda$CDM model in Fig. \ref{fig:clust_weight}. The application of $w_{it}$ to the sample reduces the clustering amplitude at the large scales as expected, since this regions is mostly affected by systematics. This reduction is significant, but inside the $1\sigma$ uncertainty limits of the fiducial correlation function. The $\Lambda$CDM model fits well till scales of $\sim130\;\mathrm{h}^{-1}\mathrm{Mpc}$, where at larger separations the clustering excess is still observed. 
     
    Even with a correcting method aggressive enough that is able to remove true power from the clustering measurements \citep{ross_2012}, the observed plateau is not significantly reduced in order to give to the full amplitude of the excess an artificial character. The goodness-of-fit of the standard model gives, $\chi^2_{red}=1.93$ for scales $3-120\;\mathrm{h}^{-1}\mathrm{Mpc}$, where it indicates a rejection of $\Lambda$CDM at the $2.5\sigma$ significance level. After including the large scales ($3-210\;\mathrm{h}^{-1}\mathrm{Mpc}$) we measure $\chi^2_{red}=1.5$, which shows a rejection of the standard model at $2\sigma$ level.

    We fit a non-Gaussian model with a linear bias as measured in Section 3.3 to the corrected clustering results, where after minimizing the $\chi^2$-statistics with the full covariance we measure the best-fit \fnl parameter, $f_{\mathrm{NL}}^{\mathrm{local}}=104\pm33$ at $68\%$ CL with $\chi^2_{red}=1.18$. The $95\%$ CL results are $46<f_{\mathrm{NL}}^{\mathrm{local}}<158$.

   The \fnl value measured from the sample after applying the $w_{it}$ weights agree with the tightly constrained results by \citet{xia_2010} $10<f_{\mathrm{NL}}^{\mathrm{local}}<106$, \citet{xia_nvss} $25<f_{\mathrm{NL}}^{\mathrm{local}}<117$ and \citet{xia_2011} $5<f_{\mathrm{NL}}^{\mathrm{local}}<84$ all in $95\%$ CL. In addition now a substantial overlap is observed between these \fnl results and the measurements found in \citet{slosar_2008}, $-29<f_{\mathrm{NL}}^{\mathrm{local}}<70$.

  \subsubsection{Correlation method}

    In this section we correct for  observational systematics by applying the method  developed by \citet{scranton_2002,ross_2011,ho2012}, where we will refer to it as the \emph{correlation} method. In this approach we calculate the auto-correlation function of quasars and each systematic, as well as the cross-correlation of the
    sample with the external sources of systematic, giving us information on the amount of correction we need to apply on the clustering results as a function of scales.

    We use the HEALPix maps to calculate the auto and cross-correlation of the quasar and the potential systematics, where for $N_{side}=64$ we create roughly pixels of size \mbox{$\sim0.84\;deg^2$}. We repeat this process for every redshift slice after splitting the sample into bins of size $\Delta z=0.03$ and applying the angular mask of the BOSS quasar. In each pixel $i$ and redshift shell we measure the overdensity of the quantity in question as
    
    \begin{equation} \label{eq:overd}
    \delta_{i,z}=\frac{x_{i,z}}{\overline{x}_z}-1
    \end{equation}
    
    \noindent where $x_{i,z}$ is the value of the quantity in pixel $i$ and redshift slice $z$ and $\overline{x}_z$ is the average of the quantity over all pixels inside the redshift slice $z$. The auto/cross-correlation functions of the quasars and systematics is calculated from
    
    \begin{equation} \label{eq:pixel_cf}
     \xi(s)=\frac{\sum_{i,j,z1,z2}\delta_{i,z1}\delta_{j,z2}\Theta_{i,j,z1,z2}(s)N_1(z1)N_2(z2)}{\sum_{i,j,z1,z2}\Theta_{i,j,z1,z2}(s)N_1(z1)N_2(z2)}
    \end{equation}
    
    \noindent where the sum is over the different pixels $i,j$ at redshift slices $z1$ and $z2$, which are the redshift bins of the pixel $i$ and $j$ respectively, $\Theta_{i,j,z1,z2}(s)$ is 1 if the separation between two pixels is within the bin $s\pm\delta s$ and 0 otherwise, $N_2(z2)$ is the number of quasars in the redshift slice $z2$. To calculate the cross-correlation of systematics with an angular map and the quasar field, one has to keep the systematic's overdensity field constant with redshift as well as assign to it a flat $n(z)$.

    The amplitude of the effect caused in the clustering of the quasars by the presence of systematics as a function of scale can be calculated by following \citet{ross_2011,ho2012}, where to the first order the overdensity field is given by
    
    \begin{equation}\label{eq:sum}
     \delta_{obs}=\delta_t+\sum_i \epsilon_i\delta_i
    \end{equation}
    
    \noindent where $\delta_{obs}$ is the observed overdensity of the quasar sample, $\delta_t$ is the true overdensity, $\delta_i$ is the overdensity of the $ith$ systematic and finally $\epsilon_i$ is the amplitude of the systematic's effect. In the case where the systematics are not separable with each other the corrected correlation function will be given by
    
    \begin{equation} \label{eq:cf_true_2}
     \xi_{true}(s)=\xi_{obs}(s)-\sum_i \epsilon_i^2\xi_i(s)-\sum_{i,j>i}2\epsilon_i\epsilon_j\xi_{i,j}(s)
    \end{equation}

    \noindent where the $\xi_i(s)$ is the auto-correlation of systematic $i$ and $\xi_{i,j}(s)$ is the auto/cross-correlation of the different potential systematics. Here we consider, as we did in the weighting method, that each systematic is separable and hence we calculate individually the effect of each one. As we showed in Section 5.3.1 there is only a small correlation between the systematic caused by the presence of stars and Galactic extinction, which will not significantly affect the correction method. Hence the cross-correlation terms between the different systematics in Eq. \ref{eq:cf_true_2} will be zero. We just need to calculate $\epsilon_i$ and subtract it from the observed correlation function separately for each individual potential systematic $i$. The corrected correlation function will finally be
    
      \begin{figure}
   \centering
   \resizebox{\hsize}{!}{\includegraphics{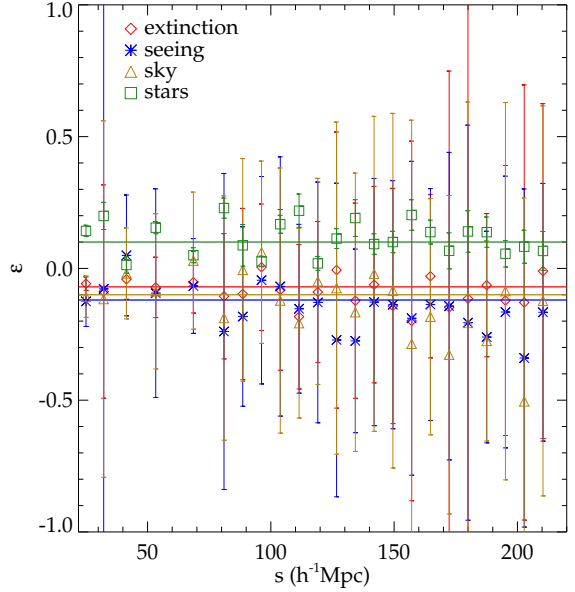}}
   \caption{The value of $\epsilon$, with the jackknife error-bars, for galactic extinction (red diamonds), atmospheric seeing (blue stars), sky brightness (yellow triangles) and star masking (green squares). The solid lines represent the best-fit constant $\epsilon$ value for each individual systematic with the corresponding colour.}
   \label{fig:epsilon}
   \end{figure}
    
    \begin{equation}\label{eq:ratio}
     \xi_{true}(s)=\xi_{obs}(s)-\epsilon_i^2\xi_i(s)
    \end{equation}

    \noindent where $\epsilon=\xi_{q,i}(s)/\xi_i(s)$, if we consider one systematic each time (see the Appendix of \citet{ross_2011}). We calculate $\epsilon$ for stellar masking, Galactic extinction in the r-band, seeing and sky brightness both in i-band.

    The resulting values of $\epsilon$, as a function of scales, for the different systematics considered here are plotted in Fig. \ref{fig:epsilon}. The correction method of \citet{ross_2011} requires $\epsilon$ to be constant, hence we use the best-fit constant values of $\epsilon_i$ (solid lines in Fig. \ref{fig:epsilon}) as calculated by the minimisation of the $\chi^2$ statistics (see Section 3.1). As we can see, in every case the calculated $\epsilon$ constant value fits well and is within the error bars, as they are calculated after propagating the jackknife errors of the auto/cross correlation functions. This suggests that there is no need for higher-order corrections in the linear relationship between the systematics and the observed galaxy clustering signal (Eq. \ref{eq:cf_true_2}) as it is discussed in \citet{ho2012}. We have to point out that after subtracting $\epsilon_i$ (Eq. \ref{eq:ratio}) to get the true correlation function, one has to propagate through the error of the estimated best-fit $\epsilon$ value to the corrected correlation function. This will increase the uncertainty limits of the true correlation function compared to those of the fiducial case, since after propagating the error of $\epsilon$ an additional term with a $\sigma_{\epsilon}$ amplitude will be present in the error propagation equation. The uncertainty in $\epsilon$ is small, hence we do not expect the increase in the error-bars of the corrected correlation function to be significantly large.
    
\begin{figure}
   \centering
   \resizebox{\hsize}{!}{\includegraphics{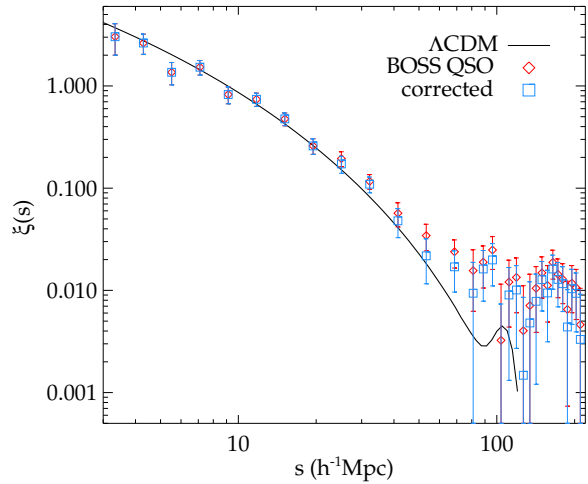}}
   \caption{The correlation function of the sample after correcting it with the sum of $\epsilon_i$ for all the systematic errors. We also present the fiducial clustering results together with the standard $\Lambda$CDM model.}
   \label{fig:ratio_cor}
   \end{figure}

    We find that the correction, after subtracting $\epsilon$ for each individual systematic, is not high enough to significantly reduce the observed excess. However if we correct the observed clustering signal for all the  potential systematics (Eq. \ref{eq:sum}), assuming they are separable (see Section $5.3.1$), then the difference between the fiducial results and the corrected ones is significant larger. The resulting correlation function together with the fiducial results and the $\Lambda$CDM model is plotted in Fig. \ref{fig:ratio_cor}.

    After fitting $\Lambda$CDM to the corrected correlation results, we measure the goodness-of-fit at scale range $3-120\;\mathrm{h}^{-1}\mathrm{Mpc}$ to find $\chi^2_{red}=1.22$. This indicates a rejection of the model
    at the $1.1\sigma$ level. The goodness-of-fit after including the large scales of clustering is $\chi^2_{red}=1.42$, which is a rejection of the standard model at the $1.8\sigma$ level. This rejection levels show that $\Lambda$CDM is an acceptable model to fit the small and intermediate scales after correcting the observed clustering signal.

                \begin{table}
    \centering
    \begin{tabular}{l|c|c}
    \hhline{===}
    Systematics & $f_{\mathrm{NL}}^{\mathrm{loc}}\;(1\sigma)$ & $\chi_{red}^2$ \\ \hline
    Extinction & $125\pm40$ & $1.14$ \\
    Seeing & $126\pm41$ & $1.16$ \\
    Sky & $126\pm40$ & $1.15$ \\
    star & $122\pm42$ & $1.18$ \\
    sum & $105\pm44$ & $1.19$ \\
    \hhline{===}
    
    \end{tabular}
    \caption{The best-fit $f_{\mathrm{NL}}$ parameter after correcting the correlation function of the sample for each systematic error separately. After considering them separable (see Section 5.3.1) we also correct the sample with their sum and present the \fnl results under the name ``sum''. The errors are calculated after allowing the bias to vary inside its uncertainty limits. The value of the reduced $\chi^2$-test from the full covariance is also presented.}
    \label{table:cor_results}
    \end{table}
    
    We fit a non-Gaussian model to the corrected clustering
    measurements, as being a better choice to model the yet persisting
    large scale excess. The correction for each systematic is not high
    enough to evidently decrease the measured non-Gaussian amplitude.
    Indeed after applying the corrections for each systematic separately
    to the quasar correlation function  we do not observe any
    significant difference with the fiducial \fnl results (see Table
    \ref{table:cor_results}). We can see that the \fnl value after
    correcting only for extinction agrees, inside the uncertainty
    limits, with the result coming after the application of the
    $A_r<0.14$ mag extinction cut to the sample (see Section 5.2). 
    
    The \fnl results after correcting the sample with the sum of $\epsilon_i$ are also presented. The new measured value is significantly reduced, since all the errors together attribute to a reduction over all scales. These constraints are the main results of this method and they are in agreement with the measurements coming after correcting the sample with the weights method, where in both cases the systematics are assumed separable.

    Usually the correlation method requires smaller size of pixels (e.g. $N_{side}=256$) than that used here. The reason is that it measures the correlation function from the overdensity in each pixel (or cubes in the 3D case), which is calculated from the objects inside them. If the pixel size is big then the measured overdensity will average out information of the internal object's clustering. Since the correlation function in this method is calculated from the pixels and not the objects (see Eq. \ref{eq:pixel_cf}), this method can lead to a 'smoothed' clustering measurements for large pixels. Hence to avoid the loss of clustering information one has to choose small enough pixels, where at the same time they must include a significant number of objects able to define an overdensity for the quantity in question. 
    
    Here we choose almost a square degree pixel and a redshift slice of $\Delta z=0.03$. The reason for this choice is that the BOSS CORE quasar sample after applying the redshift and completeness cuts is a low number density sample, roughly $6-8\;\mathrm{qso/deg}^2$ in the masked region. In order to have enough pixels in each redshift slice with an overdensity coming from more than one object, we had to use such big pixels. Due to that we expect the $\epsilon$, calculated from the auto/cross correlation function of pixel's overdensity, to under-correct the sample from the presence of potential systematics. Hence we consider as our final results on primordial non-Gaussianity after correcting the sample from potential systematics those coming from the weights method (see Section 5.3.1), $46<f_{\mathrm{NL}}^{\mathrm{local}}<158$ at $95\%\;CL$. We present the main results of the paper in a summary table (Table \ref{table:main_results}).

    \subsubsection{Cross-correlation test}
    
    Recently a novel method was developed \citep{agarwal_2013} in order to account for unknown systematic sources in a sample.  The cross-correlation of two LSS samples that have no redshift overlap should be negligibly small, after neglecting lensing effects that can introduce a magnification bias for high redshift objects.  However in a real dataset the cross-correlation signal may be non-zero due to a redshift overlap or systematics \citep{pullen_2012}. Cross-correlating two samples, corrected for the known systematic fields and with no redshift overlap between them, can give an insight of the leftover unknown systematics. In \citet{agarwal_2013} they use cross-correlations across redshift slices of the same dataset to estimate an ``unknown contamination coefficient'', which in fact does not correct for any unknown sources but rather than estimate their contribution and indicate which scale bins are dominated by unknown systematics leaving only those with a non-significant contamination. 
    
       \begin{figure}
   \centering
   \resizebox{\hsize}{!}{\includegraphics{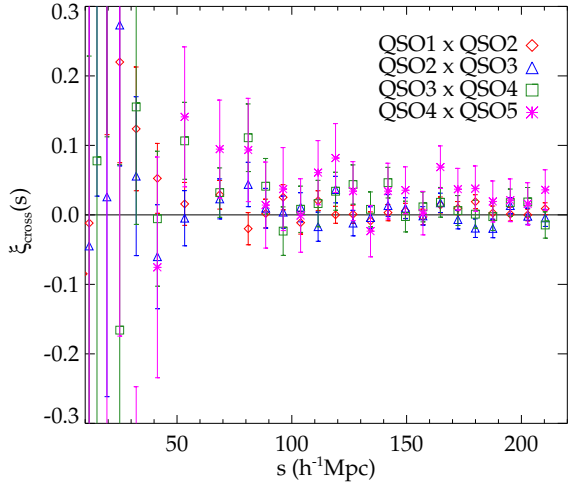}}
   \caption{The cross-correlation functions across the $5$ redshift slices of the the weighted quasar sample. The remaining cross-correlations from the rest of the combinations between redshift bins are not included, since the redshift width of the slices is larger than the maximum scale bin used to measure the auto/cross- correlation functions. The errors are from the jackknife resampling.}
   \label{fig:cross}
   \end{figure}
    
    This method uses pixelised overdensity fields in order to calculate the auto/cross-correlations for the systematics and  the objects, as the correlation method described in the previous section. As we have discussed before (Section 5.3.2), a pixelised overdensity field of the CORE BOSS quasars needs large pixels due to its low number density. This choice smooths the overdensity field and information is lost at small scales. Hence in this section we will not follow \citet{agarwal_2013}, but we will just cross-correlate the quasars across redshift slices with no overlap in order to gain information on the presence of any unknown systematics that can significantly affect our clustering and mainly the large scales. We do not take into account any magnification effects, since they have a small impact on the large linear scales of clustering \citep{scranton_2005,mcquinn_2013,deputter_2014} which are the main interest of this work. We should, though, except some cross-power due to magnification in the small and intermediate scales.  
    
    We divide the weighted quasar sample of Section 5.3.1 into 5 non-overlapping redshift bins (labelled QSO1 through QSO5), $z=2.2-2.34, 2.34-2.48, 2.48-2.62, 2.62-2.76, 2.76-2.9$ with the number of quasars in each bin being $7049, 6533, 4405, 2598$ and $1773$ respectively. To calculate the cross-correlation between the different redshift slices we use the estimator by \citet{guo_2012}, which is just a modified version of the Landy $\&$ Szalay estimator (Eq. \ref{eq:L&S}) for two different samples
    
    \begin{equation}
     \xi_{12}(s)=\frac{N_{R1}N_{R2}}{N_{D1}N_{D2}}\frac{D1D2(s)}{R1R2(s)}-\frac{N_{R1}}{N_{D1}}\frac{D1R2(s)}{R1R2(s)}-\frac{N_{R2}}{N_{D2}}\frac{D2R1(s)}{R1R2(s)}
    \end{equation}
    
    \noindent , where the subscripts $1$ and $2$ indicate the contribution to the pairs from each cross-correlating slice. $N_{R}, N_{D}$ are the randoms and quasars number respectively in each redshift slice as in the Landy $\&$ Szalay estimator. Randoms follow the same redshift cuts to construct the slices as the data. In Fig. \ref{fig:cross} we present the results of  cross-correlating across the redshift slices mentioned above.

   As we can see, the cross-correlations across redshift are non-zero for the small and intermediate scales up to $\sim 80\;\rm{h^{-1}Mpc}$. This result is expected due to intrinsic clustering and less importantly because we have not accounted for any magnification effects that may be present.  At larger separations ($\gtrsim90\;\rm{h^{-1}Mpc}$) the cross-correlation drops and is consistent with a zero-signal inside the uncertainty limits (see Fig. \ref{fig:cross}). We observe a small increase in the large scales in the cross-correlation between the last two redshift slices, but again this is consistent with zero inside the error limits. These variations around zero are partially due to the fact that the redshift slices are not fully non-overlapping, since objects that are close to the redshift cuts can introduce cross-power across the redshift slices.

   The method of cross-correlations across redshift slices, as a way to detect any present unknown systematics in the sample, is more clear-cut in the case of an angular clustering study (as in \citet{agarwal_2013}) than in the 3D case as used in this work. Hence, non zero cross-correlation signals as those observed in Fig. \ref{fig:cross} have some chance of being real, even if the cross-power comes from non-adjacent redshift slices. This implies that a non zero cross-correlation  is insufficient by itself to indicate the presence of unknown systematics in the weighted sample. In fact the main  issue  may be whether the correlations within and across redshift slices are consistent as they appear to be on comparison of the results in Fig. \ref{fig:cross} with those in Fig. \ref{fig:clust_weight}.  
   
   To get an estimate of the cross-power amplitude, we calculate the weighted average of all the cross-correlation points in Fig. \ref{fig:cross}, where we measure \mbox{$\overline{\xi}_{cross}(s)=0.0068\pm0.002$}. This value indicates that there is no evidence of any unknown systematic sources left in the weighted sample that can affect the clustering results, since the amplitude of cross-correlations across redshift slices is smaller than the average correlation function including within slice contributions.

\section{SUMMARY AND CONCLUSIONS}
   
       \begin{table*}
    \centering
    \begin{tabular}{l|c|c|c|c}
    \hhline{=====}
    Analysis & Redshift & Number of QSOs & $f_{\mathrm{NL}}^{\mathrm{loc}}$ ($95\%$ CL) & $\chi_{red}^2$ \\ \hline
    FIDUCIAL & $2.2<z<2.9$ & $22,361$ & $70<f_{\mathrm{NL}}^{\mathrm{loc}}<190$ & $1.15$ \\
    EXTCUT($A_r<0.14$) & $2.2<z<2.9$ & $16,351$ & $77<f_{\mathrm{NL}}^{\mathrm{loc}}<170$ & $0.9$ \\
    SUM CORRECTED & $2.2<z<2.9$ & $22,361$ & $31<f_{\mathrm{NL}}^{\mathrm{loc}}<169$ & $1.19$ \\
    \bf{WEIGHTS CORRECTED} & $\mathbf{2.2<z<2.9}$ & $\mathbf{22,361}$ & $\mathbf{46<f_{\mathrm{\bf{NL}}}^{\mathrm{\bf{local}}}<158}$ & $\mathbf{1.18}$ \\
    \hhline{=====}
    
    \end{tabular}
    \caption{The summary of the best-fit \fnl measurements with their jackknife (JK) errors at $95\%$ confidence level and the $\chi_{red}^2$ values for different analyses: FIDUCIAL (the original sample with no corrections and further cuts besides those described in Section 2.2), EXTCUT (the sample after applying the $A_r<0.14$ extinction cut), SUM CORRECTED (the fiducial sample after correcting it from all the potential systematics considered here with the application of the correlation method presented in Section 5.3.2), WEIGHTS CORRECTED (the fiducial sample after applying the weights method of Section 5.3.1 for correcting potential systematics). }
    \label{table:main_results}
    \end{table*}
   
    We have measured the two point correlation function of the BOSS CORE quasars sample \citep{white2012} in the redshift range $2.2<z<2.9$. Quasars are highly biased tracers of the underlying dark matter, making them good candidates to constrain primordial non-Gaussianity. Deviation from the Gaussian initial conditions in the primordial perturbation field introduces an extra scale dependent term in the bias relationship between galaxy and matter distribution. Such a scale dependence can be more  easily detected in the large scale clustering of high biased tracers like quasars, since the signature of an  extra term in the bias would be more evident.
    
    We fit a generated standard $\Lambda CDM$ model to our clustering results at scales, $3<s<50\;\mathrm{h}^{-1}\mathrm{Mpc}$,  in order to calculate the best-fit Gaussian bias. We find that the Eulerian
    linear bias of the BOSS quasars $b_G=3.74\pm0.12$, which is in agreement with the results of \citet{white2012}. The standard model is consistent with our results out to scales of $50\;h^{-1}Mpc$, where at larger separations we observe a plateau in the amplitude of the clustering which is not compatible with the predictions of the generated $\Lambda CDM$ model. More precisely we use the full covariance matrix to measure the goodness-of-fit of the standard
    model. The data for the scale range of $3-120\;\mathrm{h}^{-1}\mathrm{Mpc}$ reject the model at the
     $2.2\sigma$ significance level. Including the large scales ($3-210\;\mathrm{h}^{-1}\mathrm{Mpc}$) $\Lambda$CDM is rejected at the $2.7\sigma$ level. These results indicate that the standard cosmological model cannot be excluded from our dataset at a high significance, not even after the inclusion of the clustering plateau, due to the high uncertainty of the large scales.
    
    We generate a model incorporating the non-Gaussian scale dependent bias and fit it to the correlation function of quasars in order to measure the amplitude of the deviation from Gaussianity in the primordial density field. We measure for the local regime, $70<f_{\mathrm{NL}}^{\mathrm{local}}<190$ at $95\%$ CL, where we treat the measured Gaussian bias as a free fitting parameter allowing it
    to vary inside its uncertainty limits in order to calculate the error of the best-fit \fnl. These results are consistent with the less constrained measured \fnl values coming from other LSS surveys like those in \citet{padmanabhan2007,ross_aj2012,nikoloudakis2012}, as well as having a small overlap with the tighter measurements of
    \citet{slosar_2008,xia_2011}.
    
    One of the biggest disadvantages of measuring the amplitude of non-Gaussianities from LSS surveys is that systematic effects that may be present can affect the clustering results and especially the large scales. Detecting and correcting for such error sources can directly affect the value and constraints of primordial non-Gaussianity amplitude as it is measured from the sample's clustering. 
    
    In order to check the sample for the presence of systematic errors we divided the sample according to the value of different observational parameters, galactic hemisphere as well as removing regions of high star density. We did not observe any significant difference between the fiducial results and those after the application of cuts, other than when we divided the sample into North and South Galactic Cap. Quasars of the South Galactic hemisphere have an apparent higher clustering amplitude at large scales than the Northern hemisphere's. The number of South hemisphere quasars is small and hence do not affect significantly the clustering of the full sample.
    
    To test the sample further we measured the relationship of four potential error sources (i.e. Galactic extinction, seeing, sky brightness and stars) with the number of quasars, where we found a significant correlation with all of them besides the systematic caused by the presence of stars. The relationship between the number of stars and quasars does not show any evident trend that could justify that the masking caused by foreground stars can significantly affect the measured quasar clustering. For the observational systematics we observe a large anti-correlation between them and the number density of quasars, where the last drops as the value of extinction, seeing and sky brightness increases. Galactic extinction has the highest correlation with the number density of quasars as seen in Fig. \ref{fig:histo_err}. To test this systematic source more thoroughly we applied three different extinction cuts, where we remove regions of the sample with extinction $A_r>0.2$, $A_r>0.16$ and $A_r>0.14$ mag. The 
highest reduction in the quasar clustering is observed for the stricter cut of $A_r>0.14$ mag. We use this sample to fit $\Lambda$CDM, where the goodness-of-fit gives a rejection at the significance level of $2.1\sigma$ and $1.7\sigma$ for the scales of $3-120\;\mathrm{h}^{-1}\mathrm{Mpc}$ and $3-210\;\mathrm{h}^{-1}\mathrm{Mpc}$ respectively. The fitted non-Gaussian model gives a best-fit \fnl of $77<f_{\mathrm{NL}}^{\mathrm{local}}<170$ at $95\%$ CL. These reduced constraints have now a significant overlap with the measurements of \citet{xia_2010,xia_nvss,xia_2011}.
    
    Trying to correct for these systematics we follow the two robust methods of \citet{ross_2011,ho2012} and \citet{ross_2011,ross_2012}. The weights method reduces the amplitude of the large scale plateau giving constraints on non-Gaussianity, $46<f_{\mathrm{NL}}^{\mathrm{local}}<158$ at $95\%$. These measurements are now overlapping with the tight results of \citet{slosar_2008,xia_2010,xia_nvss,xia_2011}. The correlation method correcting each systematic separately gave values of \fnl very close to the original indicating that neither of these systematics is responsible for the clustering excess. Assuming that these systematic are separable between each other we correct the sample with the sum of $\epsilon_i$ over all the error sources considered here. The main constraints on non-Gaussianity from this method are , $31<f_{\mathrm{NL}}^{\mathrm{local}}<169$ at $95\%$ CL. The \fnl results from the two correction methods agree inside their $1\sigma$ uncertainty limits. However we believe that the large 
size of pixels used here in order to calculate the auto- and cross-correlation of the quasar sample and systematics will eventually under-correct the effect of potential systematic, since it smooths out the clustering of the object inside the pixels. Hence the final constraints on non-Gaussianity after applying a method to correct for the systematics present in the BOSS quasars sample is the result coming after the application of the weights method. The standard cosmological model is not excluded at a high significance from the corrected datasets, where the $\Lambda$CDM rejection is $2.3\sigma$ and $1.8\sigma$ for the weights and correlation method respectively. After taking also into consideration all the previous rejection values we can conclude that the standard cosmological model cannot be rejected by the clustering results of the BOSS quasars at a high significance level. $\Lambda$CDM fits well for most of the scale range but it is an inadequate model to fit the observed large scale excess that is still persisting even after the correction of the sample for potential systematic errors. Finally we test the corrected sample for any unknown systematic sources present by cutting the dataset into 5 redshift slices and calculating the cross-correlations between slices. \citet{agarwal_2013} used cross-correlations of this kind to characterize unknown systematics. Here we just calculate the cross-power of the sample after correcting for systematics with the weights method in order to get an insight into any remaining unknown systematic sources that can affect the clustering. The cross-correlation across redshift is mostly non-zero at small and intermediate scales, as expected because of intrinsic clustering and lensing effects. At large separations the cross-power does not reveal any significant non-zero signal. The weighted average of the cross-correlations across redshift is smaller than the average correlation function including within redshift slice contributions, indicating no evidence for any further systematics in the previously corrected sample.

    The main \fnl measurements from the clustering of the BOSS quasar sample are significantly non-zero in contrast to the recent results from Planck \citep{planck_2013_xxiv}, where they calculate the amplitude of non-Gaussianity to be consistent with the predictions of the standard model ($f_{\mathrm{NL}}^{\mathrm{local}}=2.7\pm5.8$
    at $68\%$ CL). However, in \citet{planck_2013_xv} they found that the low-$l$ spectrum of the Planck data ($l\lesssim30$) deviates from the best-fit $\Lambda$CDM model at the $2.7\sigma$ significance level. This could have essential implications for the parameters estimated by Planck including $f_{\mathrm{NL}}$. In addition to this, in \citet{planck_2013_xxiii} they found a significant deviation from Gaussianity in the form of positive
    kurtosis of the wavelet coefficients, which is in contrast to the measured amount of non-Gaussianity from the CMB angular bispectrum (see Tables 2,3,4 of \citet{planck_2013_xxiii}). These findings are more likely to correspond to numerous anomalies (e.g. dipolar power modulation, hemisphere asymmetry, generalised power modulation, phase correlations) observed at the large angular scales of the Planck sky. Most of these features  were also detected in the WMAP data, ruling out the possibility that they are systematic artefacts. More tests have been made in \citet{planck_2013_xxiii} proving that they are real features of the CMB. The nature of the anomalies is unknown, and the polarization data to be released in 2014 are expected to give the information needed to resolve this issue. 
    
    The 2-point correlation function of the SDSS-III BOSS CORE quasars sample suggests that there are either additional systematics that we did not account for or spurious fluctuations or \fnl is not zero as predicted by the standard model. If these two cases are excluded after a more detailed analysis of the data, then for a $f_{\mathrm{NL}}^{\mathrm{local}}\sim0$ we need to find different theoretical models that can produce a scale dependent bias other than primordial non-Gaussianity. Improved tests for non-Gaussianity will soon be possible, since BOSS continues to measure quasar redshifts that will increase the size and completeness of the quasar sample.

\section*{ACKNOWLEDGEMENTS}

We thank A.D Myers for his work on constructing the BOSS mask used here, as well as N. Nikoloudakis and A.J. Ross for their useful help and comments on this work.

Funding for SDSS-III has been provided by the Alfred P. Sloan Foundation, the Participating Institutions, the National Science Foundation, and the U.S. Department of Energy. The SDSS-III web site is http://www.sdss3.org/. 

SDSS-III is managed by the Astrophysical Research Consortium for the Participating Institutions of the SDSS-III Collaboration including the University of Arizona, the Brazilian Participation Group, Brookhaven National Laboratory, University of Cambridge, University of Florida, the French Participation Group, the German Participation Group, the Instituto de Astrofisica de Canarias, the Michigan State/Notre Dame/JINA Participation Group, Johns Hopkins University, Lawrence Berkeley National Laboratory, Max Planck Institute for Astrophysics, New Mexico State University, New York University, Ohio State University, Pennsylvania State University, University of Portsmouth, Princeton University, the Spanish Participation Group, University of Tokyo, University of Utah, Vanderbilt University, University of Virginia, University of Washington, and Yale University.


\bibliographystyle{mn2e}
\bibliography{bibliographypaper}

\end{document}